\documentclass[trackchanges, preprint2, twocolumn]{aastex6}

\usepackage{natbib}

\usepackage{braket}
\usepackage{graphicx}
\usepackage{amsmath}
\usepackage{amssymb}
\usepackage{physics}

\DeclareMathOperator*{\argmax}{arg\,max}

\newcommand{\avg}[1]{\left\langle #1 \right\rangle}

\begin{document}

\title{Inference of Unresolved Point Sources At High Galactic Latitudes \\ Using Probabilistic Catalogs}

\author{Tansu Daylan}
\affiliation{Department of Physics, Harvard University, Cambridge, MA}
\author{Stephen K. N. Portillo}
\affiliation{Harvard-Smithsonian Center for Astrophysics, Cambridge, MA}
\author{Douglas P. Finkbeiner}
\affiliation{Department of Physics, Harvard University, Cambridge, MA}
\affiliation{Harvard-Smithsonian Center for Astrophysics, Cambridge, MA}

\date{July 18, 2016}

\begin{abstract}
Detection of point sources in images is a fundamental operation in astrophysics, and is crucial for constraining population models of the underlying point sources or characterizing the background emission. Standard techniques fall short in the crowded-field limit, losing sensitivity to faint sources and failing to track their covariance with close neighbors.  We construct a Bayesian framework to perform inference of faint or overlapping point sources. The method involves probabilistic cataloging, where samples are taken from the posterior probability distribution of catalogs consistent with an observed photon count map. In order to validate our method we sample random catalogs of the gamma-ray sky in the direction of the North Galactic Pole (NGP) by binning the data in energy and Point Spread Function (PSF) classes. Using three energy bins spanning $0.3 - 1$, $1 - 3$ and $3 - 10$ GeV, we identify $270\substack{+30 \\ -10}$ point sources inside a $40^\circ \times 40^\circ$ region around the NGP above our point-source inclusion limit of $3 \times 10^{-11}$/cm$^2$/s/sr/GeV at the $1-3$ GeV energy bin. Modeling the flux distribution as a power law, we infer the slope to be $-1.92\substack{+0.07 \\ -0.05}$ and estimate the contribution of point sources to the total emission as $18\substack{+2 \\ -2}$\%. These uncertainties in the flux distribution are fully marginalized over the number as well as the spatial and spectral properties of the unresolved point sources. This marginalization allows a robust test of whether the apparently isotropic emission in an image is due to unresolved point sources or of truly diffuse origin.
\end{abstract}

\maketitle

\section{Introduction}
\label{sec:introduction}

Inference of the underlying point sources in photon count maps is a recurring problem in astronomy. Potential challenges include poorly known backgrounds, detector noise, shot noise from faint or modeling degeneracies from overlapping point sources. The resulting symptom is that of flux incompleteness, where the faintest sources are not resolved, but instead absorbed into the diffuse background prediction. Hence the flux distribution of an incomplete catalog exhibits an unphysical roll off at the faint end, even when there are yet fainter true point sources in the image.

A commonly applied approach to point source inference is the frequentist method of asking whether an additional point source increases the maximum likelihood compared to the null model, i.e., without the point source. Iteratively performing this exercise over the image while potentially floating other parameters such as the background normalization, one can produce a map of delta log-likelihood, which can be used to identify features in the count data that are unlikely to come from a diffuse emission component. A standard in gamma-ray astronomy is to reject model point sources that yield a test statistic below 25, which, for a $\chi^2$ distribution with four degrees of freedom (spatial coordinates, flux and spectral index), corresponds to just above 4$\sigma$. This approach is computationally cheap and has been the standard algorithm to reduce full-sky maps to catalogs such as the 3FGL catalog of point sources in the gamma-ray sky \citep{Acero2015a}. In the limit where an image is populated by well-separated point sources, this method can capture the spatial and spectral uncertainties of individual sources in the form of ellipsoids assuming Gaussian covariance. However as sources start to overlap, covariances between positions and spectra of model point sources cannot be captured by whether it is favorable to reject the null hypothesis for a single source. Instead, there can be multiple sources consistent with the observed data with a potentially complicated spatial and spectral covariance.

Generalization of the frequentist approach to pairs and even multiplets of point sources can in principle probe such covariances between point source positions and fluxes. However the computational complexity quickly increases before a significant fraction of the parameter space volume can be explored. Also, the best fit delta log-likelihood comparison of models with different numbers of point sources becomes ill-defined since a point source model can fit the data at least as well as another with fewer sources. Therefore one needs to balance the goodness of fit of a point source model with its predictivity. If this type of across-model comparison is not properly handled, the maximum likelihood catalog will either blend point sources or over-fit the image by introducing spurious point sources.

Traditional (deterministic) catalogs can reduce large amounts of observations to relatively concise lists of point sources, precluding false positives with the use of a hard significance cut that discards subthreshold information. However it is important to keep in mind that catalogs are still models that describe our state of knowledge consistent with the given data up to statistical and systematic errors. Especially in the crowded field limit, the uncertainties on the number, localizations and spectra of candidate point sources are more complicated than the usually adopted Gaussian form. Therefore keeping a fair ensemble of realizations of the underlying catalog space properly propagates the uncertainties to subsequent analyses that rely on the catalog. This is in contrast with the frequentist approach of representing our state of knowledge about the point sources with an estimator of the most likely catalog. These concerns make a case for adopting a Bayesian approach to point source inference.

In this paper we construct a Bayesian framework, Probabilistic Cataloger (\texttt{PCAT}), to perform probabilistic point source inference. In this setting, the hypothesis space is the union of emission models that have a number of point sources between $N_{min}$ and $N_{max}$, along with parameters characterizing the diffuse emission and the PSF. Therefore the number of point sources in a member model, $N$, itself becomes a discrete parameter of the \emph{metamodel}, i.e., top-level model. We then sample from the posterior of the metamodel given the observed data by implementing the necessary transdimensional proposals to jump across models. This precludes the necessity to run separate MCMC simulations for each model in order to estimate the Bayesian evidence, which is subject to large uncertainties. When marginalized over all other parameters, the posterior distribution of the model indicator, i.e., the number of point sources in the model, can be used to calculate the relative evidence (the Bayes factor) between models. Given that detailed balance is ensured at each across-model proposal, models with too many point sources not justified by the data are less frequently visited, since most of the added parameter space is wasted, i.e., inconsistent with the data. Therefore the resulting Bayes factor penalizes models for the loss of predictivity as well as goodness of fit.

Our work inherits elements from and builds on probabilistic cataloging discussed in \citep{Hogg2011}. The resulting statistical model is hierarchical in the sense that the hierarchical priors we place on the point source parameters, e.g., positions, fluxes and colors, are parametrized by a small set of hyperparameters, which, in turn, admit hyperpriors. During the evolution of the MCMC state, the sampler can also propose to change the hyperparameters by respecting the imposed prior. Because they parametrize the prior distribution of the point source parameters, the posterior distribution of the hyperparameters encode our state of knowledge about the population characteristics. This allows one to statistically probe the source count function below the detection threshold of traditional catalogs. 

In general, there are two distinct questions that can be posed about the underlying distribution of point sources in a given image:
\begin{itemize}
    \item What is the flux distribution of the most significant $N$ point sources?
    \item What is the number of point sources above a given minimum flux, $f_{min}$?
\end{itemize}
For small values of $f_{min}$, traditional catalogs, by construction, can only address the first question, whereas a probabilistic catalog can provide an answer to both questions after proper marginalization of the posterior samples. In principle, it is also possible to float the hyperparameter, $f_{min}$, during the generation of a probabilistic catalog. However, in the limit of arbitrarily small $f_{min}$, diffuse origin hypothesis is nearly degenerate with that of a population of unresolved point sources. Therefore, our hierarchical model becomes insensitive to changes in $f_{min}$, when $f_{min}$ becomes much smaller than the typical fluctuations in the image. We pay particular attention to how we choose $f_{min}$, which will be discussed in Appendix \ref{sect:info}.

Another approach to probing the source count function at the faint end is the fluctuation analysis where the 1-point probability distribution function of the emission is used to estimate the contribution of unresolved point sources to the total emission \citep{Scheuer1957}. By modeling the tail of the distribution of deviations due to unresolved point sources just below the detection threshold, fluctuation analysis can distinguish truly isotropic emission from unresolved point sources. The method has been considered across the whole electromagnetic spectrum, e.g., in radio \citep{Vernstrom2014, Condon2012}, Far Infrared (FIR) \citep{Friedmann2004}, x-rays \citep{Fabian1992, Worsley2004} and gamma-rays \citep{Faucher-Giguere2010, Zechlin2015, Lee2014}.

After introducing the Bayesian framework, we use our formalism to construct a probabilistic catalog of the gamma-ray sky in the North Galactic Polar Cap (NGPC). We then compare our results with the traditional 3FGL catalog published by the Fermi-LAT Collaboration and show that our median catalog agrees with the 3FGL as well as revealing low-significance point sources. Nevertheless, the real benefit of constructing a probabilistic catalog becomes apparent in crowded regions such as the inner regions of the Milky Way. Given the intriguing possibility that the inner galaxy GeV excess could be due to unresolved point sources \citep{Lee2016, Bartels2016}, we leave the application of probabilistic cataloging to the inner galaxy data, to a dedicated future work.

The rest of the paper is structured as follows. We begin in Section \ref{sect:modl} by introducing our emission model, leading into a discussion of our hierarchical inference framework. Then, in Section \ref{sect:samp} we present the sampling method we use for probabilistic cataloging. We present our results on mock data and the NGPC region in Sections \ref{sect:mock} and \ref{sect:ngal}, respectively. We then provide a discussion of our results in Section \ref{sect:disc} and conclude in Section \ref{sect:conc}.

\section{The hierarchical Bayesian model}
\label{sect:modl}
\subsection{Modeling photon emission}
The emission from a point source can be modeled as a delta function in position space with some parametrized energy spectrum, convolved with the spatial and spectral instrument response, i.e., the point and line spread functions, of the measuring instrument, respectively. Given that the Fermi-LAT energy resolution of $\sim 10\%$ is smaller than our energy bin width of $\sim 100\%$, we assume infinite energy resolution. The delta function at the position of each point source gets convolved with the PSF, $\dd{\mathcal{F}}/\dd{\Omega}$, in units of the fraction of total flux per solid angle. This yields the model point source flux map, $\mathcal{P}_{im}$, in energy bin $i$ and PSF class $m$, with units of 1/cm$^2$/s/sr/GeV, when summed over all point sources.
\begin{align}
    \mathcal{P}_{im} (l, b) = \sum_{n=1}^N & \dv{\mathcal{F}_{im}}{\Omega} (l_n, b_n) f_{ni}
\end{align}
Here, $f_{ni}$ is the flux of the $n^{th}$ point source in the $i^{th}$ energy bin. We assume a power law for the spectral energy distribution so that
\begin{align}
    f_{ni} = f_n \Bigg(\frac{E_i}{E_0}\Bigg)^{-s_n},
\end{align}
where $E_0=1.7$ GeV is the pivot energy, $f_n$ and $s_n$ are the normalization and power law index of the spectral flux of each point source. At $\sim$ GeV energies, simply modeling the spectra of the point sources using a power law is not an accurate description of the data. Galactic gamma-ray emitters such as pulsars are known to exhibit an exponential cutoff in their spectra at $\sim 1-10$ GeV. Nevertheless, since our Region of Interest (ROI) in this work is restricted to the NGPC, most of the sources are expected to be extragalactic. Moreover, extragalactic gamma-ray sources such as blazars and radio galaxies have curved spectra, i.e., a power law with a running index, due to their broadband inverse Compton emission on the external radiation field of the Active Galactic Nucleus (AGN). To address this concern we restrict the energy span to 1.5 decades, i.e., between 0.3 GeV and 10 GeV, where curved spectra can be approximated using a single power law. We do not attempt the full probabilistic regeneration of the 3FGL, which uses data between 100 MeV and 100 GeV by relying on a spectral model with a larger number of free parameters.

For a given ROI, there may also be emission from diffuse or extended sources. We therefore include into our model the diffuse emission prediction provided by the Fermi-LAT Collaboration \citep{Acero2016}, which accounts for Inverse Compton Scattering (ICS) emission due to upscattering of star light by relativistic electrons as well as gamma-rays from pion decay and bremsstrahlung due to the interaction of cosmic rays with gas and dust. At high galactic latitudes such as in the NGPC, where the latter process dominates, the model is roughly proportional to the ISM column density \citep{Ackermann2011} as traced by dust \citep{Schlegel1998}.

In addition to the spatially varying diffuse emission model, we also add an isotropic template into our background prediction. This component serves two purposes. First, it models isotropic emission, whether of cosmic or instrumental origin. Second, it absorbs potential emission from point sources below the inclusion flux limit. With the addition of the background emission, the total model becomes
\begin{align}
\mathcal{M}_{im} = \mathcal{P}_{im} + \mathcal{D}_i + \mathcal{I}_i.
\end{align}
We sample $\mathcal{M}_{im}$ over a \texttt{HealPix} grid of resolution $n_{side} = 256$. The normalization of both templates are allowed to float in all energy and PSF classes and admit log-uniform priors. Ideally, the diffuse model prediction, $\mathcal{D}_i$, should be smoothed to match the PSF kernel of the data in each PSF class, $m$. We did not need to perform this operation, however, given the fact that the diffuse model does not have bright features in the NGPC.

In order to be able to marginalize over uncertainties in the background prediction, we allow the normalizations of the background templates to float in each energy bin. We will denote the normalization parameters with the parameter vector $\vec{A}$ in what follows.

Finally, in a photon counting experiment, the likelihood function is the Poisson probability of observing $k_{ijm}$ counts given a mean of $\mu_{ijm}$ counts in the $i^{th}$ energy bin, $j^{th}$ pixel and $m^{th}$ PSF class. We then sum the log-likelihood over all pixels, energy bins and PSF classes to construct a joint log-likelihood across different energy bands and PSF classes.

\subsection{Modeling the population characteristics of point sources}
In general, the set of prior beliefs about the statistical behavior of a model parameter, $\theta$, can be encapsulated in its prior probability distribution, $P(\theta)$. However, in this work we express the prior on the model parameters using a hierarchical structure, which requires a distinction between different levels of prior belief. By using the word \emph{prior}, we refer to the first level in the hierarchy. This includes the prior belief that all model point source fluxes are drawn from a power-law with index $\alpha$ and that the Poisson mean of the number of point sources, $N$, is $\mu$. For this work on the NGPC, we assume that the point source positions are drawn from the uniform distribution and that the colors have a Gaussian distribution. We further assume that the model point sources have vanishing n-point spatial or spectral correlations and are independent and identically distributed realizations of an underlying population.

The prior probability distribution of the number of point sources, $N$, is given by
\begin{equation}
     P(N|\mu) = \frac{\mu^N}{N!}e^{-\mu}.
     \label{equa:numbpnts}
\end{equation}

The hyperparameter $\mu$ is taken to be log-uniform distributed such that
\begin{equation}
    P(\mu) = \dfrac{1}{\ln \mu_{max} - \ln \mu_{min}} \dfrac{1}{\mu}
\end{equation}
for $N_{min} < \mu < N_{max}$ and vanishes otherwise. This choice yields a scale-free prior on the number of point sources.

Note that we use the same notation to refer to probability densities of continuous variables, e.g., $P(\mu)$, and probabilities of discrete variables, e.g., $P(N)$. When the parameter is discrete, the notation implies probability, whereas it refers to the probability per differential interval in the parameter, if the parameter is continuous.

\explain{In our model, there are both continuous parameters and a discrete parameter. Therefore, when referring to the probability densities of continuous variables in the first draft, we used the notation $dP/da$ for some continuous random variable $a$. This was intended to emphasize the fact that we refer to probability per differential interval of parameter $a$. However, we did not maintain this convention throughout the paper to limit notational clutter and sometimes used the simpler $P(a)$ notation, which may be confusing. In this revision, we always use the latter and clearly mention in the text that we refer to probability densities when the parameter is continuous and to probabilities when the parameter is discrete.}

Furthermore, we assume that the flux of the $n^{th}$ point source in the pivot energy bin 1 - 3 GeV, $f_n$, is distributed as a power law between some $f_{min}$ and $f_{max}$ with the index $\alpha$ at the central flux bin of $\sim 10^{-9}$/cm$^2$/s/GeV.
\begin{align}
    P(f_n|\alpha) = 
    \begin{cases}
    \dfrac{1-\alpha}{f_{max}^{1-\alpha} - f_{min}^{1-\alpha}} f_n^{-\alpha} \text{ for } f_{min} < f_n < f_{max} \\
    0 \text{ otherwise}
    \end{cases}.
     \label{equa:flux}
\end{align}

We place a uniform prior on the angle described by this slope, which yields
\begin{equation}
    P(\alpha) = \dfrac{1}{\tan^{-1}(\alpha_{max}) - \tan^{-1}(\alpha_{min})}\dfrac{1}{1+\alpha^2}
\end{equation}
for $\alpha_{min} < \alpha < \alpha_{max}$ and 0 otherwise.

\subsection{PSF modeling}
\label{sect:psfn}
A particle detector inevitably introduces errors when the arrival direction of a photon event is reconstructed. At small angular deviations, the random scatter around the true arrival direction can be modeled as a Gaussian
\begin{align}
    \mathcal{G}(\theta_0; \sigma) &= \frac{1}{\sqrt{2\pi \sigma^2}} \exp\Bigg(- \frac{\theta^2_0}{2\sigma^2}\Bigg),
\end{align}
where $\theta_0$ is the angular distance from the true direction. The variance of the Gaussian is the sum of variances due to the resolution of the silicon strips and multiple scattering, which itself scales with energy. At large angular deviations, however, the Fermi-LAT PSF instead follows a power-law. A convenient function that approximates a Gaussian at small deviations while converging to a power law at large deviations is the King function
\begin{align}
    \mathcal{K}(\theta_0; \sigma, \gamma) &= \frac{1}{\sqrt{2\pi \sigma^2}} \Bigg(1-\frac{1}{\gamma}\Bigg) \Bigg(1 + \frac{\theta_0^2}{2\sigma^2\gamma}\Bigg)^{-\gamma},
\end{align}
where the Gaussian scale is given by $\sigma$ and the power law slope is set by $\gamma$.

The PSF modeling of the Fermi-LAT Collaboration uses a weighted sum of two such functions \citep{Ackermann2013a}. Therefore, there are $5N_eN_{psf}$ independent parameters in the model, where $N_e$ and $N_{psf}$ are the number of energy bins and PSF classes. We use this model when we do not float the PSF parameters.

When none of the point sources in the ROI are bright enough, such as Geminga or Vela, to constrain the PSF tails, floating them causes large uncertainties in the inferred background and point source fluxes. This is because the tails of the PSF are nearly degenerate with the background normalization and allowing them to float without strong priors causes a significant bias in the flux predictions. We generalize the algorithm to sample from the PSF parameters, when needed, and float the PSF during our mock runs. However, we fix the PSF during the nominal data run and discuss the results of letting the PSF float in Appendix \ref{sect:psfnvari}. The ability to float the PSF is especially relevant for ground-based optical astronomy, where the PSF is different in each exposure.

In the other case, where we do float the PSF, a typical ROI without exceptionally bright point sources cannot constrain a double-King function with a floating scale factor. In particular, a bright source is needed to break the near-degeneracy between the power law slopes of the core and tail components of the radial profile. Therefore, in that case we fix the scale factor to the best-fit value provided by the Fermi-LAT Collaboration and use a linear combination of a King function and a Gaussian with only $4N_eN_{psf}$ free parameters. We place uniform priors on the logarithm of the angular scale, $\sigma$, the relative fraction of the Gaussian and King components, $f$, and the inverse tangent of the slope, $\gamma$.

\begin{equation}
    \mathcal{P} = a_G \mathcal{G}(\theta_0; \sigma_G) + a_K \mathcal{K}(\theta_0; \sigma_K, \gamma)
\end{equation}

Inference of the point source catalog sensitively depends on the PSF modeling. Due to the power-law tails of the PSF, bright members of the point source population can outshine the faintest point sources even $\sim$ a few degrees away as shown in Figure \ref{figr:eval}.

In the rest of the paper, we will collectively denote the set of parameters characterizing the PSF, with the parameter vector $\vec{\eta}$.

\subsection{Prior structure}

\begin{figure}[ht]
    \centering
    \includegraphics[width=0.45\textwidth, trim=4.cm 2cm 2.cm 1.cm, clip]{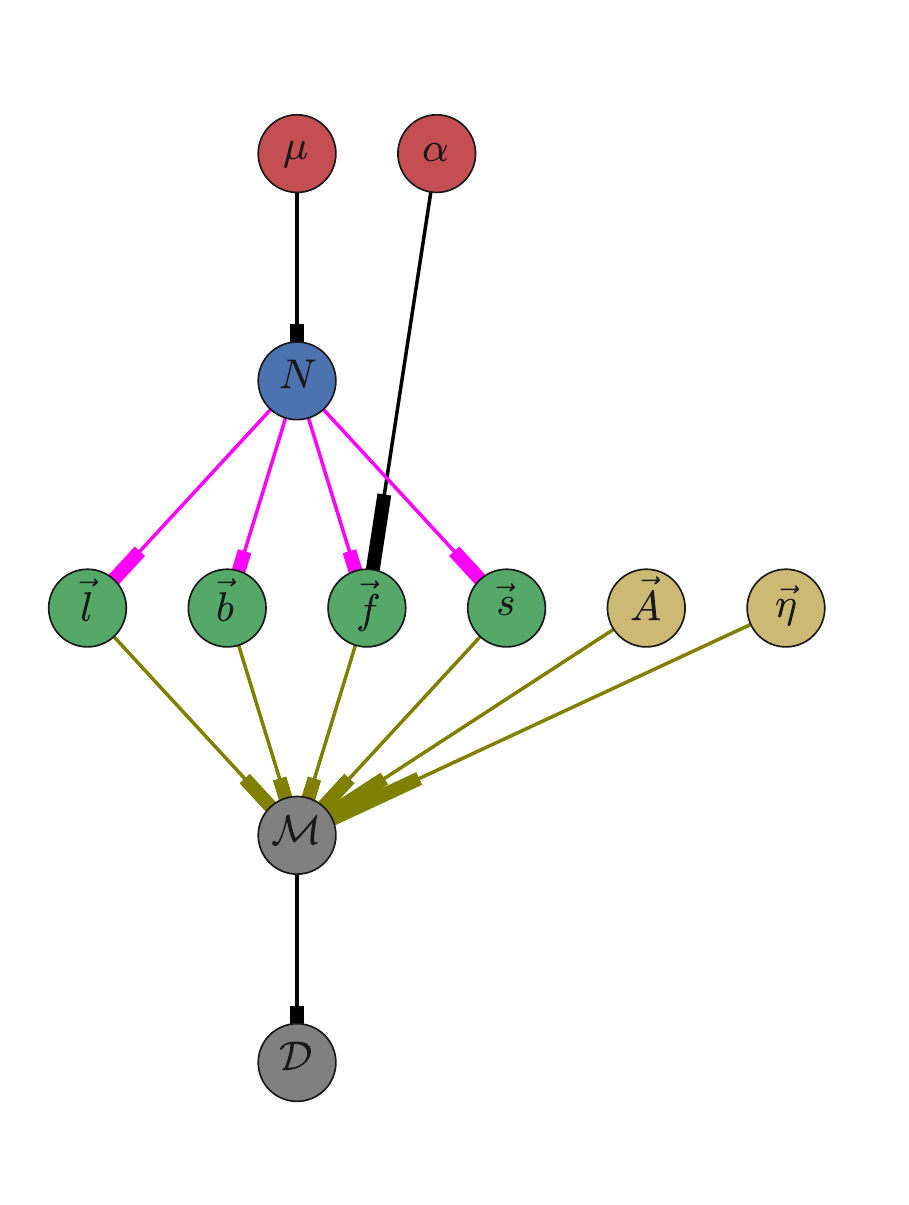}
    \caption{A Probabilistic Graphical Model (PGM) of our transdimensional model. See the text for details.}
    \label{figr:grap}
\end{figure}

A probabilistic graphical model of our inference framework is presented in Figure \ref{figr:grap}. In this representation, nodes denote random variables, while edges directed into a node denote the set of nodes that hierarchically parameterize the probability distribution of the given node, such as in Equations \ref{equa:numbpnts} and \ref{equa:flux}. The red, blue, green and yellow nodes represent our model parameters, which are assigned prior probability distributions. In particular, the red nodes are the hyperparameters $\mu$ and $\alpha$, which set the normalization and slope of the point source flux distribution, respectively. The blue node indicates the number of point sources in a model, i.e., is the multiplicity of each green node. Likewise, the green nodes are the point source parameters, i.e., longitude, latitude, flux and spectral index from left to right. The yellow nodes are the background normalizations, $\vec{A}$, and PSF parameters, $\vec{\eta}$. $\mathcal{M}$ node is a deterministic function of the above model parameters representing the set of forward modeled photon count maps. Finally, $\mathcal{D}$ represents the observed photon count maps, whose consistency with the former drives the evolution of the MCMC state through the Poisson likelihood. We also color code the directed edges such that black edges denote a probabilistic relation, whereas olive lines show a deterministic relation. Finally, magenta lines imply that the multiplicity of the destination node is set by the origin. Note that we do not use plate notation, since the multiplicity itself is a discrete parameter in our model, which admits a hierarchical prior.

Given the parameter and hyperparameters introduced above, the joint prior probability distribution of a model with $N$ point sources becomes
\begin{multline}
    P(\mu, \alpha, N, \vec{A}, \vec{\eta}, \vec{l}, \vec{b}, \vec{f}, \vec{s}) = \\
                P(\mu) P(\alpha) P(N|\mu) P(\vec{A}) P(\vec{\eta}) \prod_{n=1}^N P(l_n) P(b_n) P(f_n|\alpha) P(s_n)
\end{multline}
where we use the vector notation to refer to the union parameter set of $N$ point sources, e.g., $\vec{l} \equiv (l_1, l_2, ..., l_N)$.

\explain{This equation replaces Equation 21 in the previous draft.}

\subsection{ROI margins}

The observed count data in the ROI can potentially be affected by point sources outside the ROI. In order to model such emission from point sources just outside the ROI, we make the spatial prior region slightly larger (1 degree larger on all sides) then the ROI window over which the likelihood is calculated. Therefore the model point sources can move slightly out of the image and probe whether a feature close to the boundary can be fit better by a model point source outside the ROI. As a result the offset provides a smooth transition from a data-informed region well inside the ROI to a prior dominated region outside the ROI, where the posterior asymptotes to the prior. This can be clearly seen in the artificial accumulation of sampled point sources along the boundary of the stacked posterior, as will be shown in Section \ref{sect:ngal}.

\section{Sampling Method}
\label{sect:samp}

In this section, we describe the method used to sample from the probability distribution of catalogs consistent with the given photon count map. The starting point is the Bayesian assumption that there exists an underlying probability distribution that characterizes our knowledge of the model parameters. These parameters are the longitude, $l$, latitude, $b$, flux, $f$, and color, $s$, of each point source. Refer to Section \ref{sect:modl} for details on how we define and place priors on these parameters. The parameter space of the point source metamodel is, then, the union of the parameter spaces of the point source models that contain from $N_{min}$ up to $N_{max}$ point sources. Denoting this space as \emph{the catalog space}, $\mathcal{C}$, we therefore sample from 
\begin{equation}
    \mathcal{C} = \bigcup_{N=N_{min}}^{N_{max}} \mathcal{C}_N = \bigcup_{N=N_{min}}^{N_{max}} \mathcal{L}_N \times \mathcal{B}_N \times \mathcal{F}_N \times \mathcal{S}_N,
\end{equation}
where $\mathcal{C}_N$ is the catalog subspace with $N$ point sources. $\mathcal{C}_N$ can further be written as the product space of $\mathcal{L}_N$, $\mathcal{B}_N$, $\mathcal{F}_N$ and $\mathcal{S}_N$ that denote the longitude, latitude, flux and color spaces of point sources in the model with $N$ point sources. Although sampling from a given $\mathcal{C}_N$ can be performed by constructing an MCMC chain using ordinary Metropolis updates over $\mathcal{L}_N$, $\mathcal{B}_N$, $\mathcal{F}_N$ and $\mathcal{S}_N$, the fact that the catalog space is transdimensional means that it cannot be explored by simple within-model proposals.

We assume that the number of point sources in the image is not known a priori. Therefore, the dimensionality of the point source model, $N$, becomes a discrete parameter subject to inference. In addition to the catalog space, the complete hypothesis space includes normalizations of the isotropic and spatially varying background models in each energy bin and PSF class, as well as parameters that characterize the PSF. We will defer the discussion of these degrees of freedom to Section \ref{sect:modl}, since they have fixed dimensionality and can be explored using ordinary within-model updates. 

\subsection{Trans-dimensional sampling}

In the rest of the paper we will generically refer to a parameter in the model by $\theta$. The objective is to sample from the posterior distribution of $\theta$ given the observed count map $D$, $P(\theta|D)$, after updating our prior beliefs about $\theta$, $P(\theta)$. The Bayesian update is accomplished through the likelihood of observing the data given our model, $P(D|\theta)$. For this purpose we construct a Markov chain of states, $\{\theta\}$, whose asymptotic stationary distribution is the desired posterior distribution. Therefore we require that the chain is reversible with respect to the posterior, i.e., satisfies detailed balance condition. If the sampling space was fixed dimensional, this would imply that
\begin{multline}
    \int P(\theta|D) Q(\theta^\prime|\theta) \alpha(\theta^\prime|\theta) \dd{\theta} = \\ \int P(\theta^\prime|D) Q(\theta|\theta^\prime) \alpha(\theta|\theta^\prime) \dd{\theta^\prime},
\end{multline}
where $\theta$ and $\theta^\prime$ are the current and the proposed states, $Q(\theta^\prime|\theta)$ is the transition kernel from $\theta$ to $\theta^\prime$ and $\alpha(\theta^\prime|\theta)$ is the appropriate proposal acceptance probability, 
\begin{equation}
    \alpha(\theta^\prime|\theta) = \min \Bigg( 1, \frac{P(\theta^\prime|D)}{P(\theta|D)} \times \frac{Q(\theta|\theta^\prime)}{Q(\theta^\prime|\theta)}\Bigg).
    \label{equa:acppmths}
\end{equation}

Note that when sampling from the catalog space, $\theta$ and $\theta^\prime$ can have different dimensions. This brings up an interesting issue that the chain can no longer be reversible in the transdimensional case, since the transition is not one to one. The remedy is to draw random auxiliary variables, $u$ and $u^\prime$, using the densities $g(u)$ and $g^\prime(u^\prime)$, to match the dimensions of the initial and final states of the transition such that the mapping $(\theta, u) \leftrightarrows (\theta^\prime, u^\prime)$ is a diffeomorphism, where
\begin{align}
    H(\theta, u) &= (\theta^\prime, u^\prime) \\
    H^{-1}(\theta^\prime, u^\prime) &= (\theta, u).
\end{align}

The dimension matching,
\begin{align}
    \mathcal{D}(\theta) + \mathcal{D}(u) = \mathcal{D}(\theta^\prime) + \mathcal{D}(u^\prime),
\end{align}
where $\mathcal{D}$ denotes the dimension operator, conceals the transdimensional nature of the proposal and ensures reversibility. In this case we require
\begin{multline}
    \int P(\theta|D)g(u)\alpha(\theta^\prime|\theta) \dd{\theta}\dd{u} = \\ \int P(\theta^\prime|D)g^\prime(u^\prime)\alpha(\theta|\theta^\prime) \dd{\theta^\prime}\dd{u^\prime}.
    \label{equa:rjmcbala}
\end{multline}
The transition is accomplished through the transition kernel, $H(\theta, u) = (\theta^\prime, u^\prime)$, which replaces the probabilistic transition kernel, $Q(\theta^\prime|\theta)$, in the fixed-dimension case. For example, in the case where the proposal is to add a new point source, the auxiliary parameters, $u$, are simply the parameters describing the new point source. This method is known as the Reversible Jump MCMC (RJMCMC) \citep{Green1995, Green2008, Hastie2012}, which is a variant of MCMC that allows across-model moves in a pool of models indexed by their dimensionality. We inherit the reversible jump formalism in implementing transdimensional proposals.

Given the freedom to jump across models, there are an infinite number of ways to propose such transitions using the current state. However, only some are useful schedules to explore the catalog space. Denoting the probability of the $m^{th}$ type of proposal in state $\theta$ by $j_m(\theta)$, the ratio,
\begin{equation}
    \frac{j_m(\theta^\prime)}{j_m(\theta)},
\end{equation}
should be also included in the resulting acceptance rate in order to compensate for any bias in the proposal frequencies.

Using the new detailed balance condition, Equation \ref{equa:rjmcbala}, one finally obtains,
\begin{gather}
    \alpha(\theta^\prime|\theta) = \min(1, \alpha_0), \\
    \alpha_0 = \frac{P(D|\theta^\prime)}{P(D|\theta)}
    \frac{P(\theta^\prime)}{P(\theta)}
    \frac{j_m(\theta^\prime)}{j_m(\theta)} \frac{g(u^\prime)}{g(u)} \Bigg\lvert
    \frac{\partial(\theta^\prime, u^\prime)}{\partial(\theta, u)} \Bigg\rvert,
    \label{equa:accp}
\end{gather}
since the coordinate transformation requires that probability is conserved, i.e.,  
\begin{equation}
    \dd{\theta^\prime}\dd{u^\prime} = \Bigg\lvert
    \frac{\partial(\theta^\prime, u^\prime)}{\partial(\theta, u)} \Bigg\rvert \dd{\theta} \dd{u},
\end{equation}
for all $\theta$ and $u$.

The usual within-model proposals are recovered when both $u$ and $u^\prime$ have the same dimension. Given that the parameter space of the point sources has large covariances in crowded fields, it is likely that the sampler can get stuck in a likelihood island without being able to efficiently visit all high-likelihood regions. In order to prevent this, we take heavy-tailed Gaussian steps when proposing within-model transitions. However, large changes in parameters can also suffer from the prior ratio in Equation \ref{equa:accp}, especially when the prior is a power law such as for the fluxes of point sources. We therefore set the prior ratio to unity by transforming the parameters such that their prior distributions become uniform. This is accomplished through the use of the Cumulative Distribution Function (CDF) and its inverse, which map the parameters to the unit space, $\mathcal{U} = [0, 1]$, and back. Therefore, the actual sampling is performed in the unit space with the prior ratio set to unity by definition at the expense of inverse transforming the parameters back to their genuine values when the likelihood function is to be evaluated or a hyperparameter is updated. This excludes the number of point sources in the model, $N$, which is an integer parameter.

Conventional updates to point source positions, fluxes or colors, i.e., within-model proposals, only allow the sampler to explore one catalog subspace, $\mathcal{C}_N$. To explore the full space, $\mathcal{C}$, the sampler must propose updates that change $N$, by adding or removing sources.  We use two pairs of such updates, where $u$ and $u^\prime$ have different dimensions. The first type is the elementary operation of adding or deleting a point source to or from the current list of point sources. We denote this pair of proposals birth and death. They are the reverse proposals of each other unlike within-model updates that are manifestly reversible. Therefore both birth and death have to be present in the set of possible proposal types in order for detailed balance to be respected.

When a point source is to be added, an auxiliary vector, $u$, is drawn whose elements are distributed uniformly between 0 and 1. We then use the inverse CDF of the position, flux and color distribution functions to transform the uniformly distributed elements to random draws of the position, flux and color. These parameters determine the parameters of the point source to be added. In contrast, when a point source is to be killed, the CDF transform of its parameters define $u$.

The second pair of transdimensional proposals are splits and merges, where a point source is split into two, and two point sources are merged into one, respectively. This pair is especially important in the crowded field, where splits and merges make the exploration of the nearly degenerate regions of the parameter space more efficient. We refer the reader to Appendix \ref{sect:accpprob} for the implementation details of the move types.

\subsection{Time performance of the sampler}

Sampling from the posterior probability distribution of catalogs is a computationally demanding task. The primary reason is the large and variable number of parameters in the sample vector. For a typical MCMC run on a mock image with 200 point sources, the sample vector can have as many as $\sim$ 1000 parameters. Since position and flux changes are proposed one at a time in order to keep the acceptance ratio high, the typical autocorrelation time becomes $\sim$ 10000 MCMC steps. Furthermore the least significant features in the image require the highest number of MCMC steps for convergence, because they are sampled only slightly more frequently compared to a uniform Poisson background. Increasing the performance requirements of the sampler is the fact that hyperparameters parametrize the hierarchical priors on the point source parameters. This means that the convergence of parameters precedes that of the hyperparameters. A typical inference thus requires $\sim 10^8$ MCMC steps.

\begin{figure}[ht]
    \centering
    \includegraphics[width=0.45\textwidth]{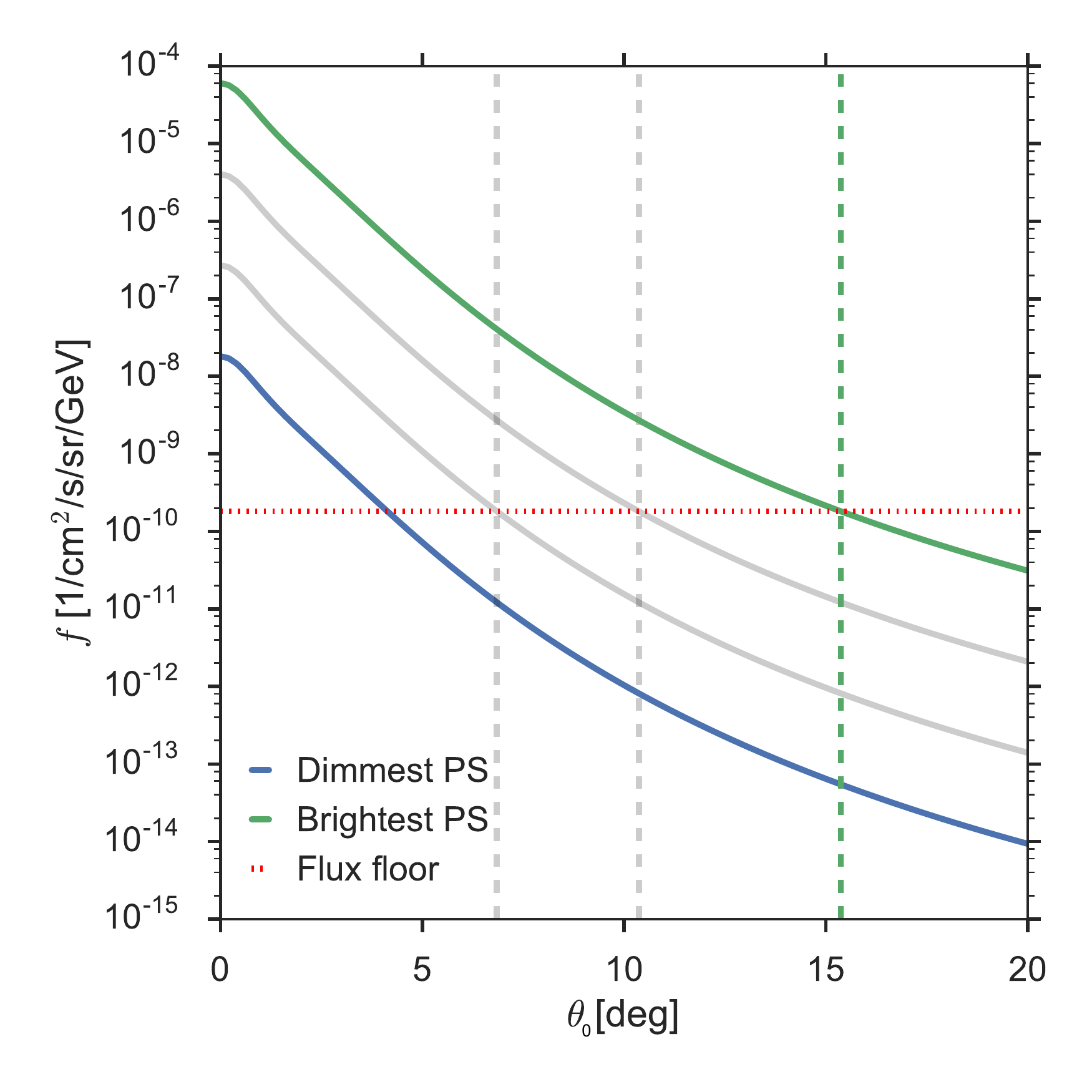}
    \caption{The radial profile of emission from a point source as reconstructed by the Fermi-LAT for different flux bins. The tail of the PSF is a power law and is not exponentially suppressed. The horizontal line highlights the minimum flux allowed by the prior multiplied by 0.01. When the position, flux or the color of a point source is changed, its contribution to the total flux map is updated only inside a circle centered at the pixel closest to the point source. The radius of the circle is set such that this approximation causes a negligible error.}
    \label{figr:eval}
\end{figure}

For each proposal in an MCMC run, the current parameter vector is used to compute the model count map, which is then compared to the data count map through the Poisson likelihood. This process typically dominates the time budget and requires careful optimization for the sampler to be scaled up to large ROIs, larger number of point sources or energy bins. We break the optimization into four steps.
\begin{itemize}
    \item Although the sampling is actually performed in the CDF-transformed parameter space, the inverse CDF transforms of the parameters are stored along with those of the CDF-transformed parameters, precluding redundant CDF transformations.
    \item Flux maps are calculated perturbatively. During each proposal the current flux map is modified by the updated, added or killed point sources. This decreases the time complexity of processing a single sample from $\mathcal{O}(N_{pix}N)$ to $\mathcal{O}(N_{pix})$, where $N$ is the number of point sources and $N_{pix}$ is the number of pixels in the ROI.
    \item For each point source, the PSF is evaluated over a subset of pixels that lie inside a circle. The list of nearby pixels for each pixel is precomputed and stored as a look-up table. The radius of the circle depends on the flux of the associated point source and is determined such that the largest flux allowed by the prior contaminates the lowest flux at most by one percent. This is illustrated by Figure \ref{figr:eval}, where the horizontal line indicates 0.01$f_{min}$, where $f_{min}$ is the minimum flux allowed by the model. Hence, point sources contribute to the total flux map up to the radius, after which the bias introduced by their neglect is significantly below the faintest possible point source. We monitor the error introduced by this approximation and ensure that the bias is negligible in all pixels, energy bins and PSF classes. As a result of this approximation the time complexity of a single proposal is further reduced to $\mathcal{O}(1)$. This implies that the time complexity of a single proposal (except those that update the PSF or hyperparameters) does not depend on how many sources or pixels there are in the image.
    \item The leading contribution to the time complexity of the average sample comes from the likelihood evaluation, which requires the computation of the proposed change to the flux over a set of data cubes, e.g., pixels, energy bins and PSF classes. In this operation, the PSF is evaluated by computing the angular distance from the point sources to the pixel centers. In order to accelerate this computation, we precompute unit vectors to the \texttt{HealPix} pixel centers, $\hat{u}_{pix}$. This allows us to compute the angular distance, $\theta_0$, using the dot product
    \begin{equation}
        \theta_0 = \arccos(\hat{u}_{PS} \cdot \hat{u}_{pix}),
    \end{equation}
    where $\hat{u}_{PS}$ is the unit vector along the point source of interest. Furthermore, we store the radial profile of the PSF in the memory and interpolate it when calculating the flux updates. In overall, the employed acceleration scheme reduces the mean computation time per sample to $\sim 6$ ms. Nevertheless, the evaluation of the PSF takes $\sim 5$ ms per sample on the average and dominates the time budget of a typical sample.
\end{itemize}
	
In order to scale the algorithm to larger ROIs, multiple energy bins or epochs, the sampler has been parallelized to run over multiple cores. The current time performance is adequate for analyzing full-sky datasets such as that of Fermi-LAT in $\sim 5000$ CPU hours. Although the execution time per sample increases linearly with the typical number of point sources in the sampler, the time it takes to get an independent sample from the catalog space scales roughly as the square of the number of point sources. This is due to the fact that as the number of parameters increases, one either has to take smaller steps in the model space, or make changes in fewer number of parameters in a given proposal, both of which increase the autocorrelation of the resulting chain. Therefore, a slight increase in the ROI size may significantly increase the convergence time. Because this scaling is not desirable for large fields or full sky analyses, it is more feasible to separately sample from the catalog space of patches that are $\sim 10$ PSFs wide. This neglects covariances between sources and background emission in different patches, but allows scaling to much larger fields.

However, further speed improvements would be needed for analyzing optical photometric data, where the typical pixel size is less than an arcsecond as opposed to being tens of arcminutes as in Fermi-LAT. In addition to the performance improvement gained by working on a Cartesian grid, this could be achieved using GPUs (Graphics Processing Unit) given the parallelizable nature of the sampling problem. 

Note that the time complexity of the overall convergence still depends on the minimum flux allowed by the model. This is because it sets the typical number of model point sources and, hence, the size of the parameter space. In a typical run with 3 energy bins, 2 PSF classes, a $40^\circ\times40^\circ$ ROI, $\sim 10^5$ pixels and $\sim 250$ model point sources, the execution time is 250 CPU hours.
    
\subsection{Convergence diagnostics}
MCMC formalism allows the exploration of complex posterior distributions with the caveat that convergence to the stationary distribution can require a long simulation time. Given that a typical catalog has thousands of parameters, the finite running time of the Markov chain might raise concerns over whether the sampled distribution is representative of the desired target distribution. One method for evaluating the chain convergence is to inspect the variance of the sampled chain. However the variance of a single MCMC chain can underestimate the true variance, since the realized chain may not have converged to the target distribution despite having a small variance. We therefore run multiple, usually around 20, chains and compare the mean of the chain variances to the variance of the means of the chains \citep{Gelman1992}, ensuring that the initial states of the chains are over-dispersed relative to the target distribution. Having $N_c$ chains and $N_s$ samples in each chain, the resulting test statistic,
\begin{equation}
    \hat{R} = \sqrt{1 + \frac{B}{W} - \frac{1}{N_s}},
\end{equation}
is known as the Potential Scale Reduction Factor (PSRF) and can be used to assess whether the chains have converged. Here, $W$ is the within-chain variance, i.e., the mean of the chain variances,
\begin{equation}
    W = \frac{1}{N_c} \sum_c^{N_c} \Bigg( \frac{1}{N_s-1}\sum_s^{N_s}(y_s - \bar{y}_c)^2 \Bigg),
\end{equation}
$B$ is the between-chain variance, i.e., the variance of the chain means,
\begin{equation}
    B = \frac{1}{N_c-1} \sum_c^{N_c}\Bigg(\bar{y}_c - \frac{1}{N_c} \sum_c^{N_c} \bar{y}_c \Bigg)^2,
\end{equation}
and the mean of the $c^{th}$ chain is
\begin{equation}
    \bar{y}_c = \frac{1}{N_s} \sum_s^{N_s} y_s.
\end{equation}
Our definition of $B$ differs from that commonly found in the literature by a factor of $N_s$, which is absorbed into the definition of $\hat{R}$. In a well-mixed chain the PSRF should be close to unity. One caveat of using the PSRF as an estimator of the convergence in this framework is that the sampled chains are transdimensional.

Note that the problem of point source inference has a labeling degeneracy. That is to say that there is an N!-fold degeneracy in the likelihood function of a point source model with $N$ point sources, since permuting the parameter labels of these $N$ point sources leaves the likelihood invariant. For any reasonably large $N$, $N!$ is larger than the number of samples that can be drawn from the posterior. Therefore, formal convergence is not possible, but also unnecessary, since well sampling only one of the degenerate likelihood peaks reveals the unique likelihood topology of the problem. 

\begin{figure}
    \centering
    \includegraphics[width=0.45\textwidth]{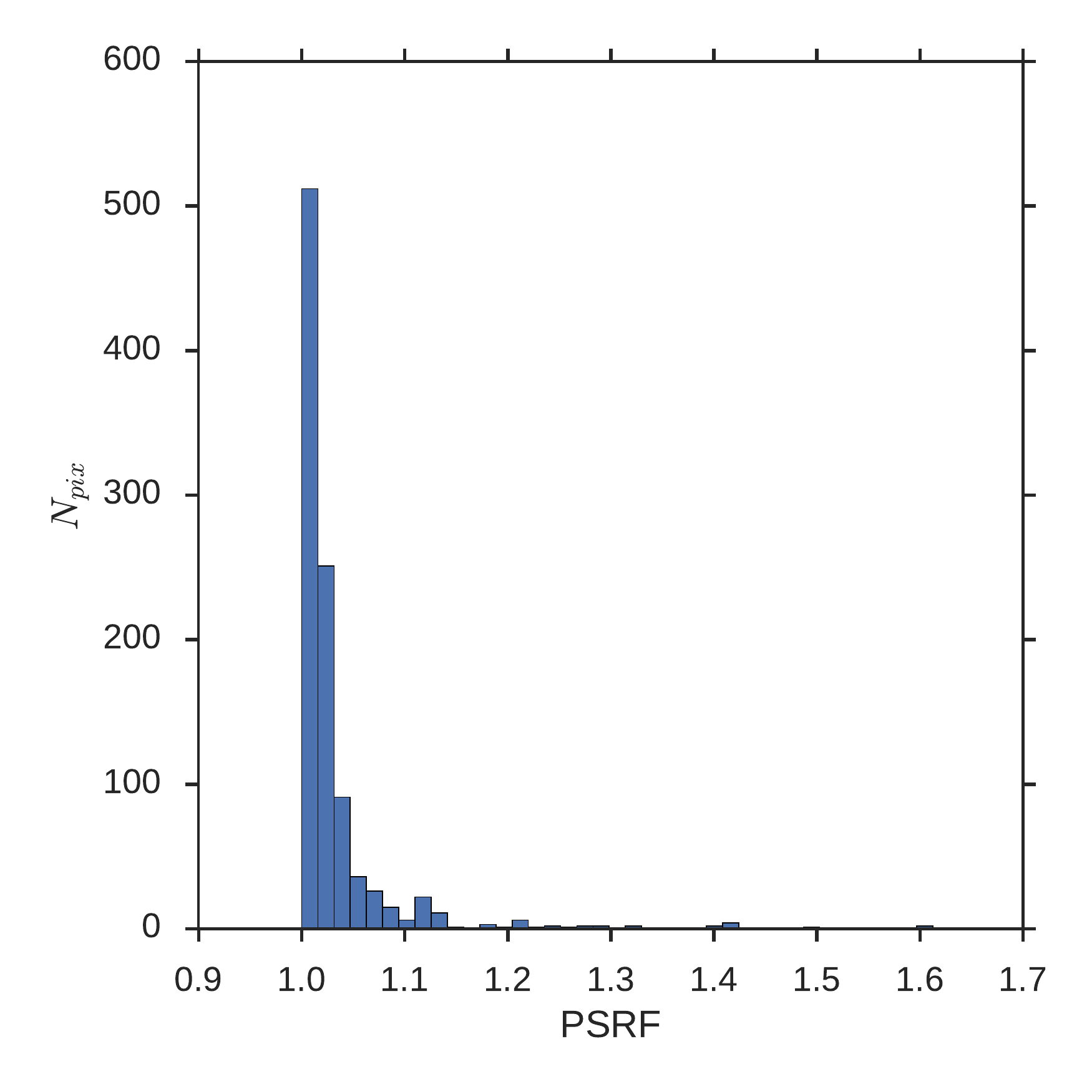}
    \caption{The distribution of the Gelman-Rubin test statistic.}
    \label{figr:gmrb}
\end{figure}

In order to probe convergence, we instead monitor the variance of the resulting model emission map. We draw 1000 random voxels (triplets of pixel, energy bin and PSF class) and show the distribution of the PSRF in our data run (Section \ref{sect:ngal}) in Figure \ref{figr:gmrb}, which confirms that the between and within chain variances are similar for most pixels.

\subsection{Autocorrelation of the chain}
Given that the sampler updates, adds or kills one point source at a time, nearby samples in the chain are expected to be similar. In order to retain the Markovian property, we therefore thin the chain by a factor equal to the maximum number of parameters allowed by the metamodel. Typically this implies that the chain is thinned by a factor of 1000-10000.

\begin{figure}
    \centering
    \includegraphics[width=0.45\textwidth]{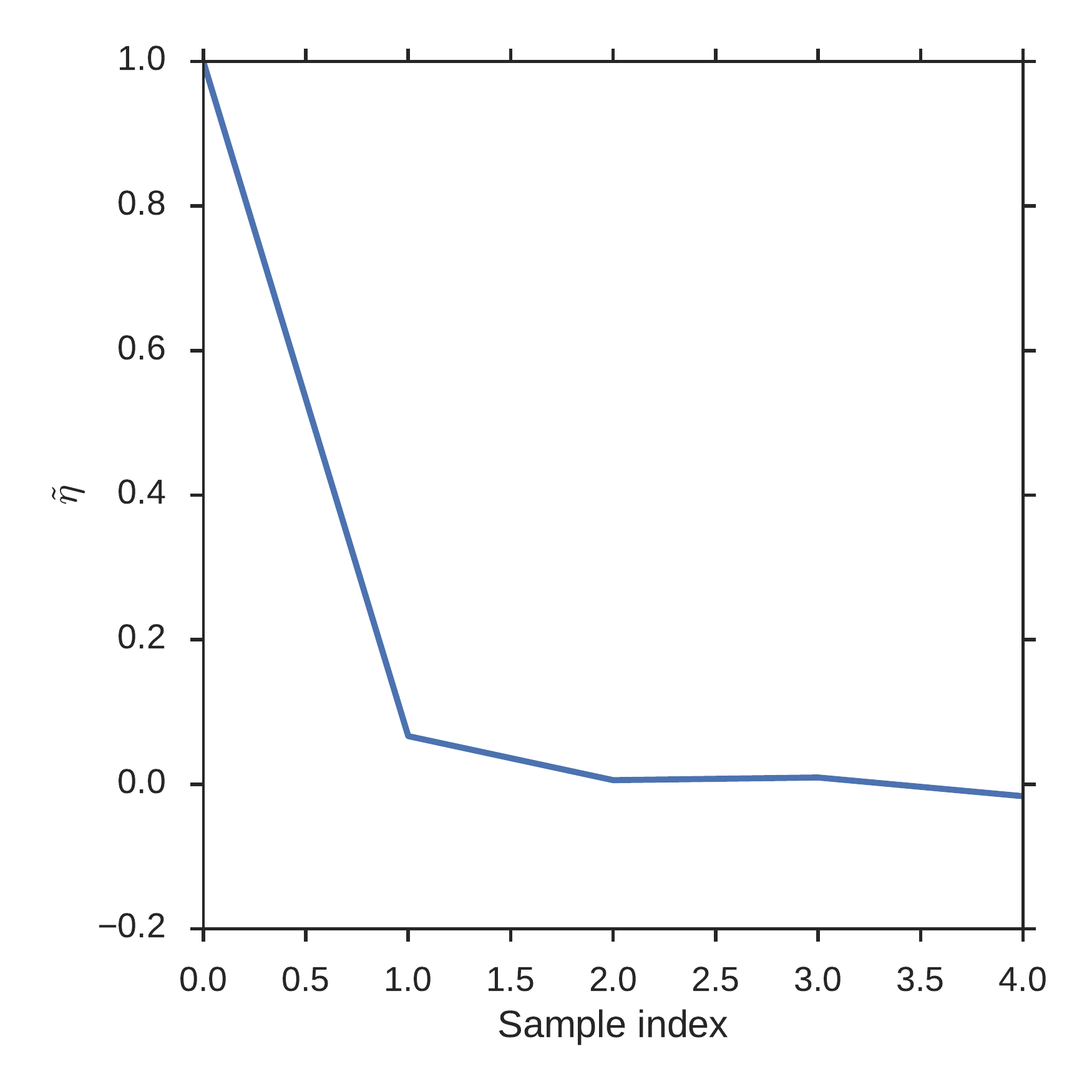}
    \caption{Autocorrelation of the diluted chain normalized by its value at zero lag.}
    \label{figr:atcr}
\end{figure}

In order to ensure that the resulting chain has the Markovian property, we compute the autocorrelation of the chain, where we follow a similar method to that of diagnosing convergence using the predicted emission map. After calculating the autocorrelation for the randomly drawn 1000 voxels, we take the average over chains and parameters. We plot the resulting autocorrelation of the chain in Figure \ref{figr:atcr}, which shows that samples in the diluted chain is memoryless.

\subsection{\texttt{PCAT}}

We make the resulting software, \texttt{PCAT}, available to the astronomy community. \texttt{PCAT} is a pure \texttt{Python 2.7} implementation of the described sampling algorithm along with extensive routines to customize the sampler for the problem at hand, further process the output, diagnose convergence and visualize probabilistic catalogs. It is designed to sample from the catalog space for a given photon count map, choice of data binning and prior structure, where the level of background and the PSF are potentially unknown. As of version 0.2, it is only intended to process binned count maps from a Poisson process.

\section{Mock runs}
\label{sect:mock}

\begin{figure*}[ht]
    \centering
    \begin{minipage}[b]{0.32\linewidth}
        \includegraphics[width=\linewidth]{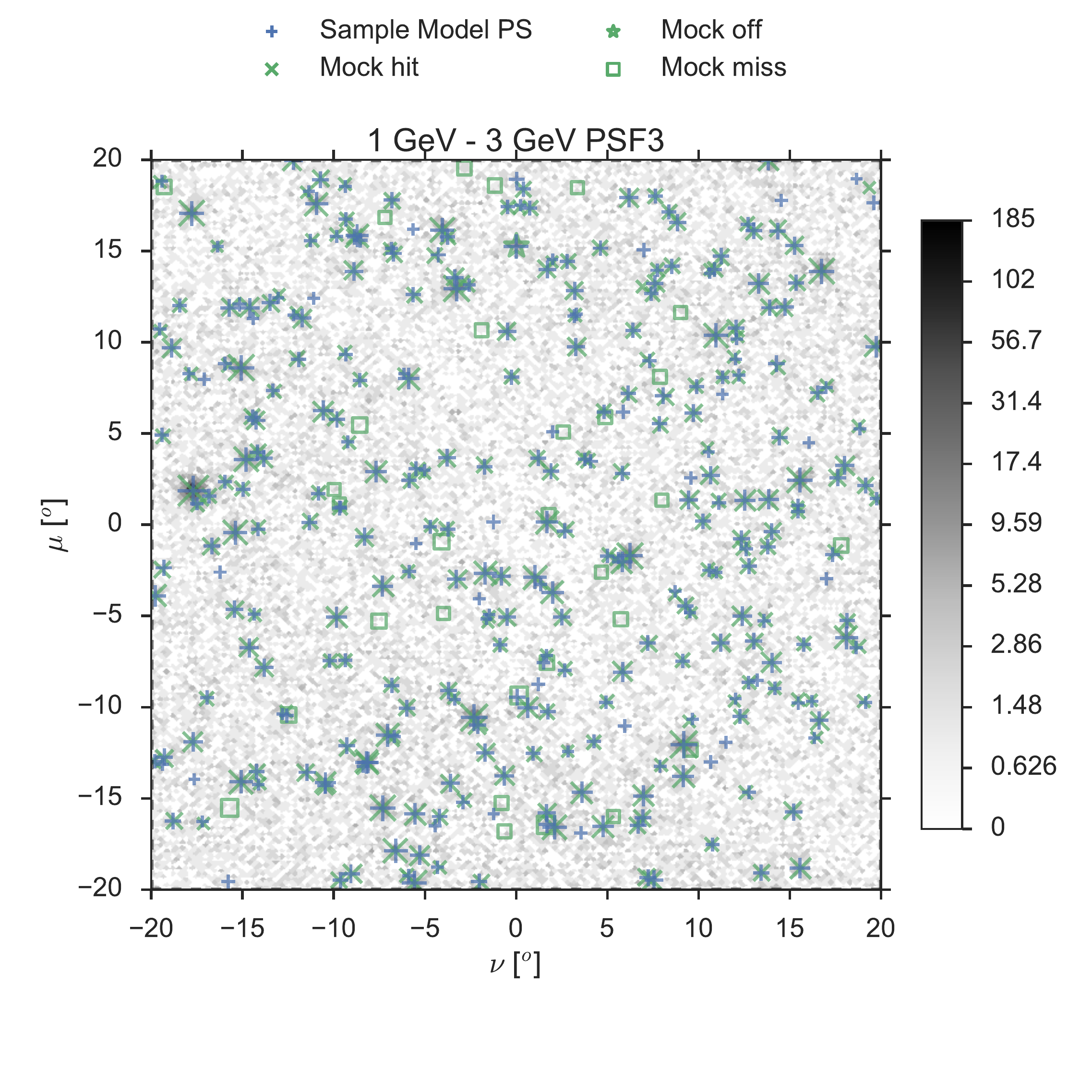}
    \end{minipage}
    \begin{minipage}[b]{0.32\linewidth}
        \includegraphics[width=\linewidth]{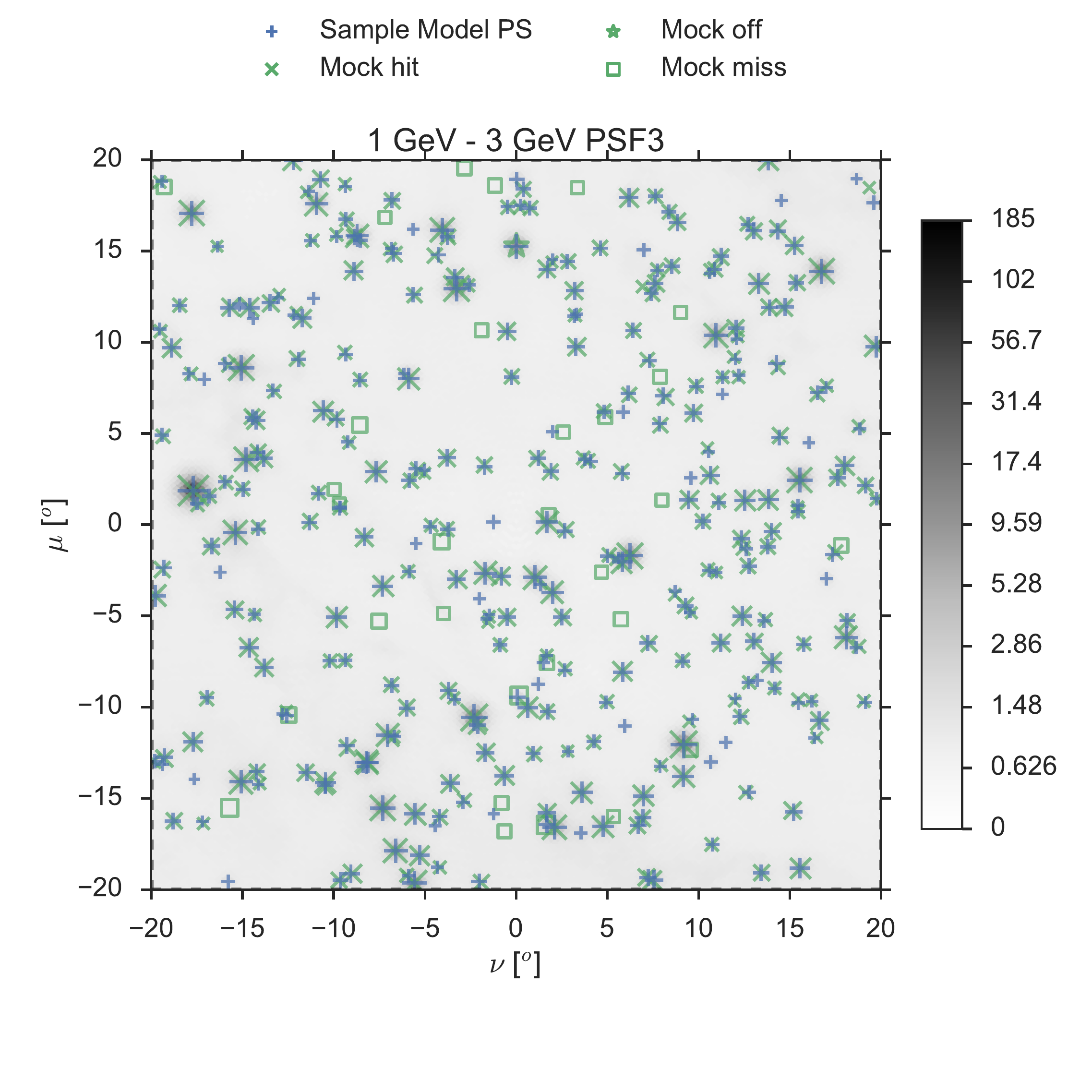}
    \end{minipage}
    \begin{minipage}[b]{0.32\linewidth}
        \includegraphics[width=\linewidth]{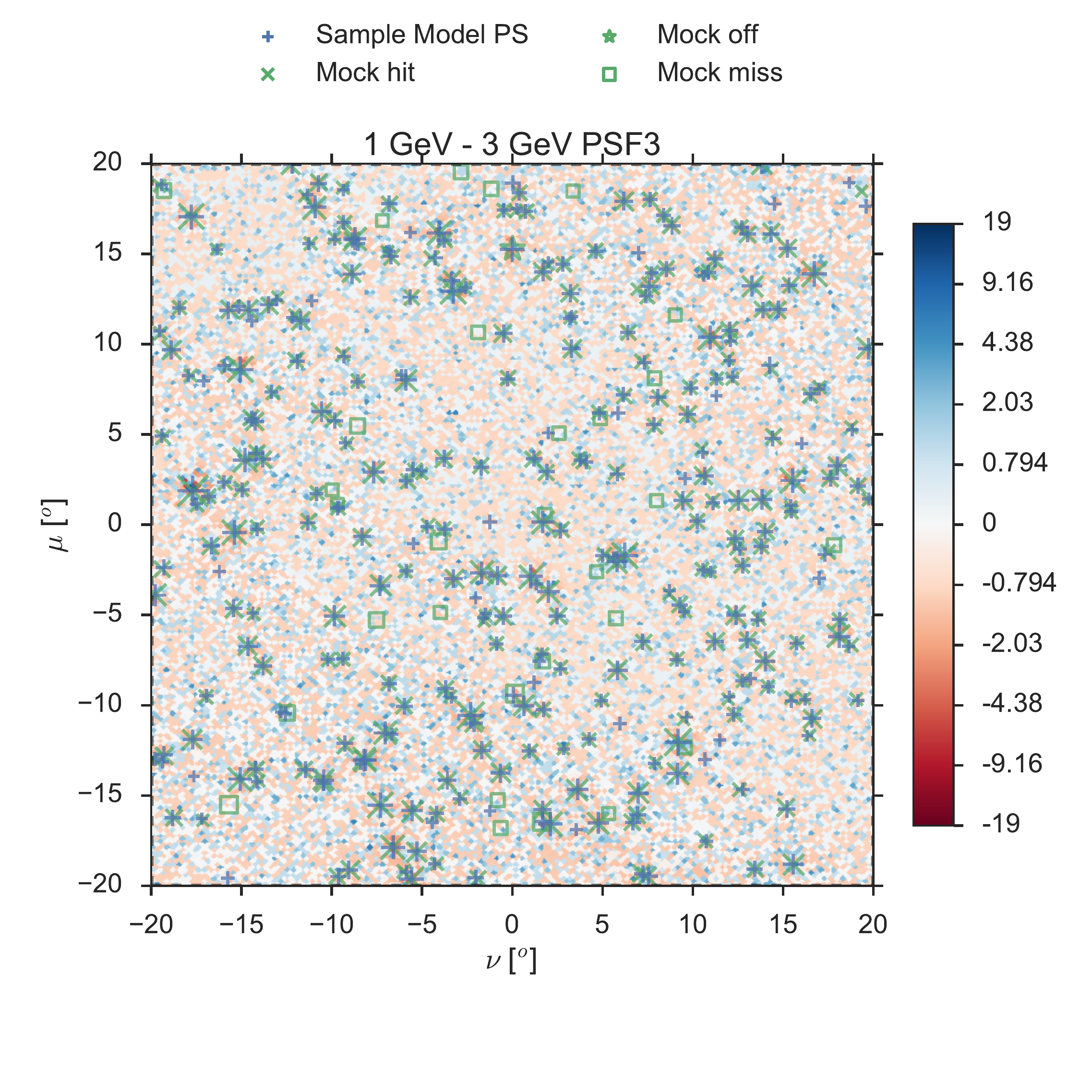}
    \end{minipage}
    \caption{A synthetic photon count map of the NGPC, where the mock data is a Poisson realization of emission due to a point source population and the diffuse background. The left and middle panels show the mock photon count map, and a sample model count map, while the right panel is the residual. The color scales indicate the number of photons per pixel in all panels and are arcsinh stretched in order to make faint features more visible. Superimposed with the count maps are the true and sample catalogs, which are shown with green Xs and blue pluses, respectively. If a true point source is missed, it is instead indicated with a green square. Likewise, if the flux of the sample does not agree with that of the true point source despite spatial association, it is shown with a green star. The sizes of the markers are proportional to the logarithm of the fluxes of the corresponding point sources.}
    \label{figr:sampcnts}
\end{figure*}

In this section we will first present an ensemble of catalogs sampled using a mock, i.e., simulated, dataset. Hence we will compare the true (input) parameter values with those obtained from the ensemble. The Poisson mean of the mock map is generated by sampling all parameters randomly from the prior and then calculating the resulting emission map due to the background emission and the point sources. Finally mock data is generated as a Poisson realization of the mean emission. In total 300 point sources are generated from a flux distribution with a power law slope set to -1.8. We use the \texttt{Pass 7}, \texttt{source} class exposure map of the Fermi-LAT instrument between weeks 9 and 217, when calculating the mock number of counts from the synthetic emission map. This ensures that the mock point sources are subject to incompleteness at the same flux as the real point sources. 

Figure \ref{figr:sampcnts} summarizes a fair sample from the probabilistic catalog. The panels show the number of counts in the generated mock map, the sampled model map and the residual, respectively from left to right. Here, the shown sample is only one of many realizations that constitute a fair draw from the underlying probability distribution. The number of counts are given in the 1 GeV - 3 GeV energy bin. As can be seen in the right panel, Poisson fluctuations near bright point sources can be large even if the data is a realization of a generative model, i.e., the model is a good description of the data. In practice, further mismodeling of the PSF can result in large, and possibly, even coherent residuals around bright point sources. Depending on how low the model point source fluxes are allowed to go, this can result in the sampling of spurious point sources around bright ones. Therefore care must be taken to employ a PSF modeling that does not bias the flux distribution of the point source population.

\begin{figure*}[ht]
    \centering
    \begin{minipage}[b]{0.9\linewidth}
        \includegraphics[width=\linewidth, trim=0 3cm 0 3cm, clip]{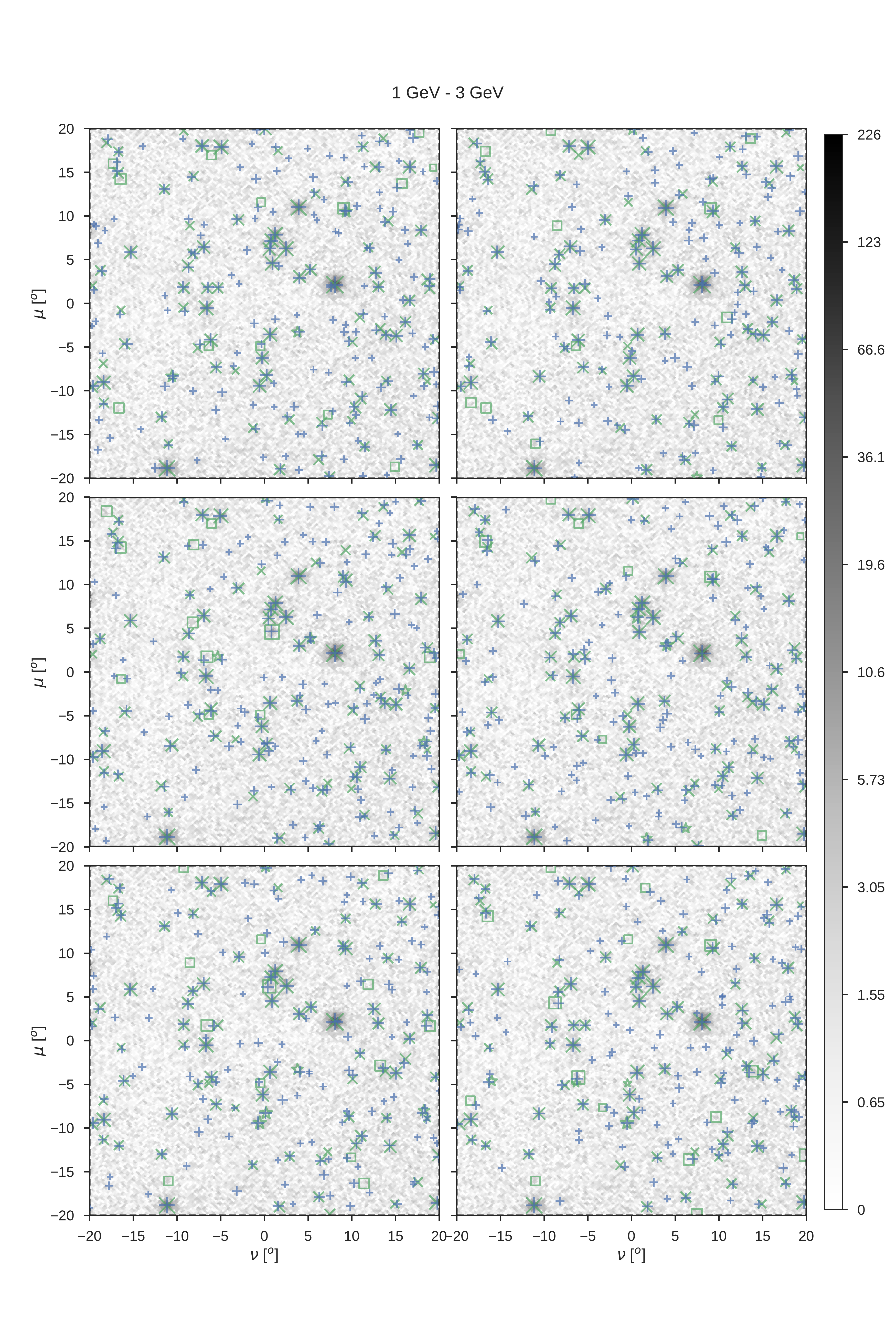}
    \end{minipage}
    \caption{The NGPC photon count map as measured by the Fermi-LAT (color scale), the 3FGL sources in the ROI (green markers) and six fair samples from the catalog space (blue pluses). The image is centered at the NGP and the axes correspond to a Cartesian projection about the NGP. The size of the markers are proportional to the logarithm of the flux of the point source. The color scale corresponding to the number of photon counts per pixel has been arcsinh stretched in order to emphasize faint features. The 3FGL sources are marked with a green square if the sample catalog does not have a model point source within 0.5 degree. Otherwise they are marked with a green X, indicating a hit. If the fluxes of the associated point sources disagree (sample flux outside the dashed lines in Figure \ref{figr:inptscatspec}), then the green X is replaced with a green star.}
    \label{figr:arrydatacnts}
\end{figure*}

Figure \ref{figr:arrydatacnts} tiles together fair samples from the posterior showing the number of data counts in each pixel, i.e., similar to the left panel in Figure \ref{figr:sampcnts}, but showing real NGPC data instead. The grid illustrates the typical evolution of the MCMC state, where bright true point sources have an associated model point source with precise spatial and spectral localization, whereas faint true point sources are only sometimes associated with model sources.

\begin{figure}[ht]
    \centering
    \includegraphics[width=0.45\textwidth]{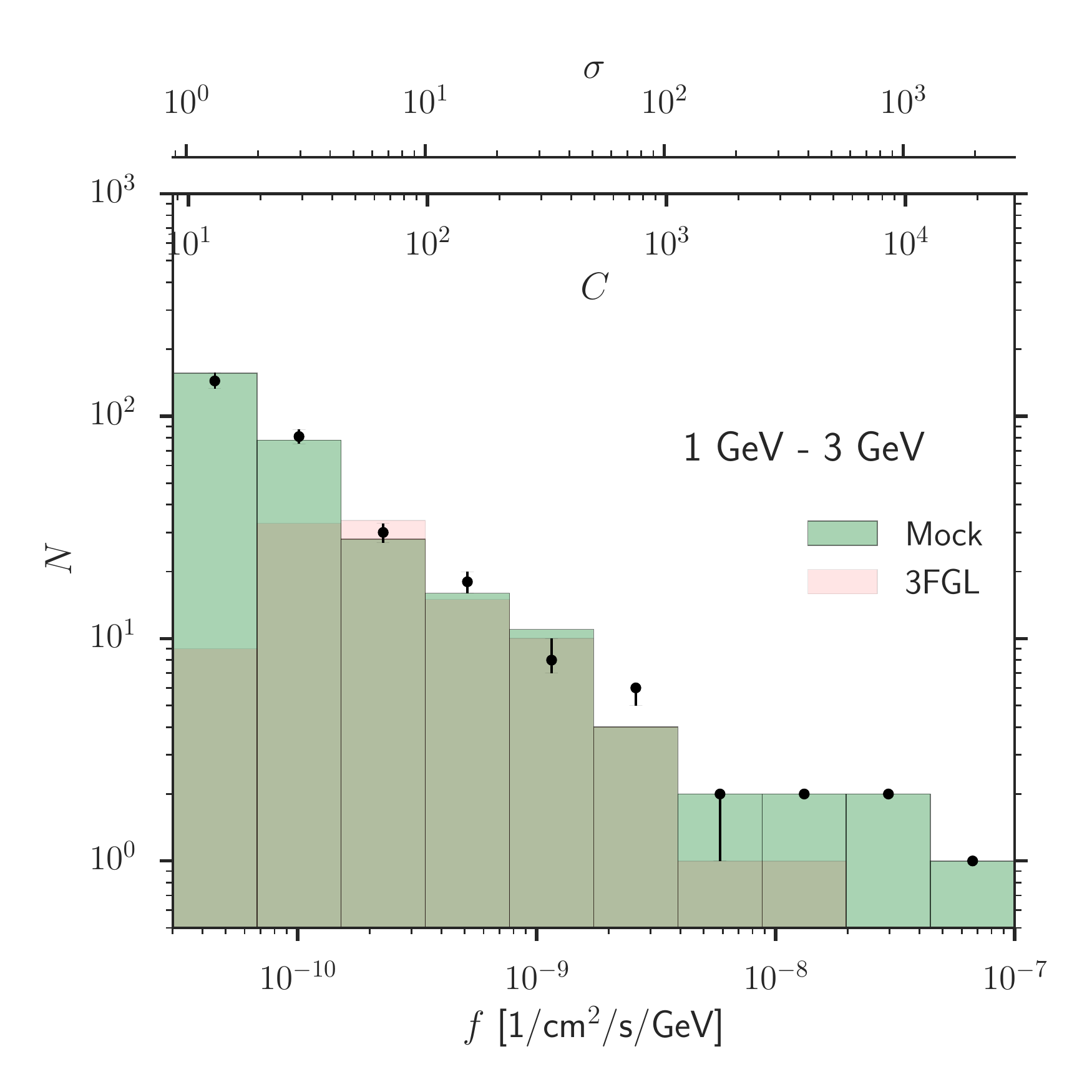}
    \caption{Posterior flux distribution function, showing its median, 68\% and 95\% quantiles. The green histogram shows the distribution of mock point sources. The red histogram highlights the 3FGL for reference. Note that the probabilistic catalog is statistically complete, i.e., has a statistically representative sample of sub-threshold point sources. The lower horizontal axis at the top shows the number of counts for the mean exposure in the ROI that corresponds to the flux axis. The top horizontal axis, then, gives the ratio of this number of counts to the number of background counts inside a Full Width at Half Maximum (FWHM), which is a proxy for the significance of a point source with the given mean number of counts in units of standard deviation.}
    \label{figr:mockhistflux}
\end{figure}

Next, Figure \ref{figr:mockhistflux} shows the flux distribution of the mock (green) and median sample (black) with the 68\% credible interval. The 3FGL is also shown with red for reference. The 1$\sigma$ uncertainties cover the truth for most of the flux bins. Note that a flux-incomplete traditional catalog rolls off at the faint end since such sources are fainter than the typical fluctuations of the background. In a probabilistic catalog the flux distribution can be probed even below where a traditional catalog would roll off. This is because even though none of the point sources in the catalog samples is to be taken as true, repeated sampling of the flux distribution can constrain the population characteristics. However, this is only true for point sources more significant than $\sim 1 \sigma$. At yet lower fluxes, the sampled flux distribution function is informed by the prior more than the likelihood, and tends to follow whatever characteristic is imposed by the hierarchical prior, i.e., a power law with a range of indices allowed by the hyperprior. This transition from a likelihood dominated region to a prior dominated one is controlled by the minimum allowed flux of the point sources, $f_{min}$, and indicates where probabilistic cataloging becomes ineffective in constraining the population characteristics. We discuss how we determine the transition region in Appendix \ref{sect:info}.

\begin{figure*}[ht]
    \centering
    \includegraphics[width=0.99\textwidth]{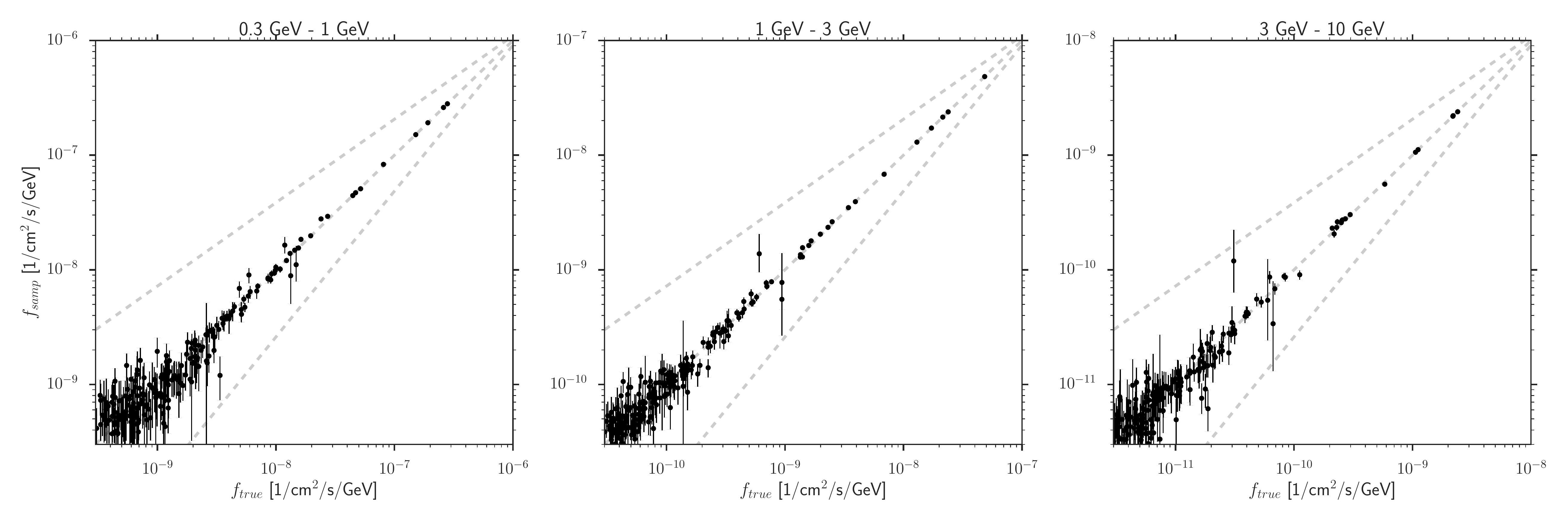}
    \caption{The median sample flux of spatial associations between the true and the probabilistic catalog. The diagonal line as well as the flux mismatch tolerance lines used to tag associations with an empty green circle in Figures \ref{figr:sampcnts} and \ref{figr:arrydatacnts}, are shown with dashed gray lines. See the text for details on matching the two. In the absence of Poisson noise or systematic uncertainties, associations would lie on the diagonal line. The vertical error bars correspond to statistical uncertainty of the probabilistic catalog and generally increase towards lower fluxes, as the observed image becomes less informative. The outliers (associations far away from a one-to-one correlation) is an artifact of associating an ensemble of sample catalogs with a traditional one in the crowded field limit, where simply associating the closest model point sources with the true catalog causes loss of information and biases the result.}
    \label{figr:mockscatspec}
\end{figure*}

The former comparison confirms the statistical agreement between the true and median catalogs. However it is also desirable to perform an element-wise comparison between the two catalogs. Towards this purpose we associate a traditional (in this case, the true) catalog and a probabilistic catalog as follows. For a given sample from the catalog space, we initialize the association algorithm by setting the fluxes of all associations to the true catalog, to zero. Then, for each point source in the sample catalog, we ask if there are any point sources in the true catalog within 0.5 degrees. If so, we take the closest such true point source and add the model point source to the true point source's list of possible associations. We then repeat this for all model point sources in the sample catalog. Along the way it happens that multiple model point sources get matched to the same true point source. In that case, we use the gathered list of possible associations to select the closest model point source as \emph{the}  association. After all samples in the probabilistic catalog has been so processed, we take the median flux associated with each point source in the traditional catalog. Figure \ref{figr:mockscatspec} illustrates the resulting correlation. The horizontal axis shows the true flux of the source, while the vertical axis marks the median of the fluxes associated to it, including potential zero fluxes due to missed associations. This procedure is generic to associations between any traditional and probabilistic catalog and not just to the association of the probabilistic catalog to the underlying true catalog.

Furthermore, in Figure \ref{figr:mockscatspec} the vertical error bars denote the statistical uncertainty due to the stacking of the ensemble of catalogs. Horizontal error bars are not provided, since the true catalog is a generated catalog without any instrumental uncertainties. At the bright end, model point sources are statistically significant and well localized, yielding small vertical error bars. Moving towards the faint end, the correlation first broadens due to associated features on the image becoming comparable to Poisson fluctuations of the background. In the extreme faint limit, one expects a given true point source to be associated with a random model point source, completely suppressing the correlation.

\begin{figure}[ht]
    \centering
    \includegraphics[width=0.5\textwidth]{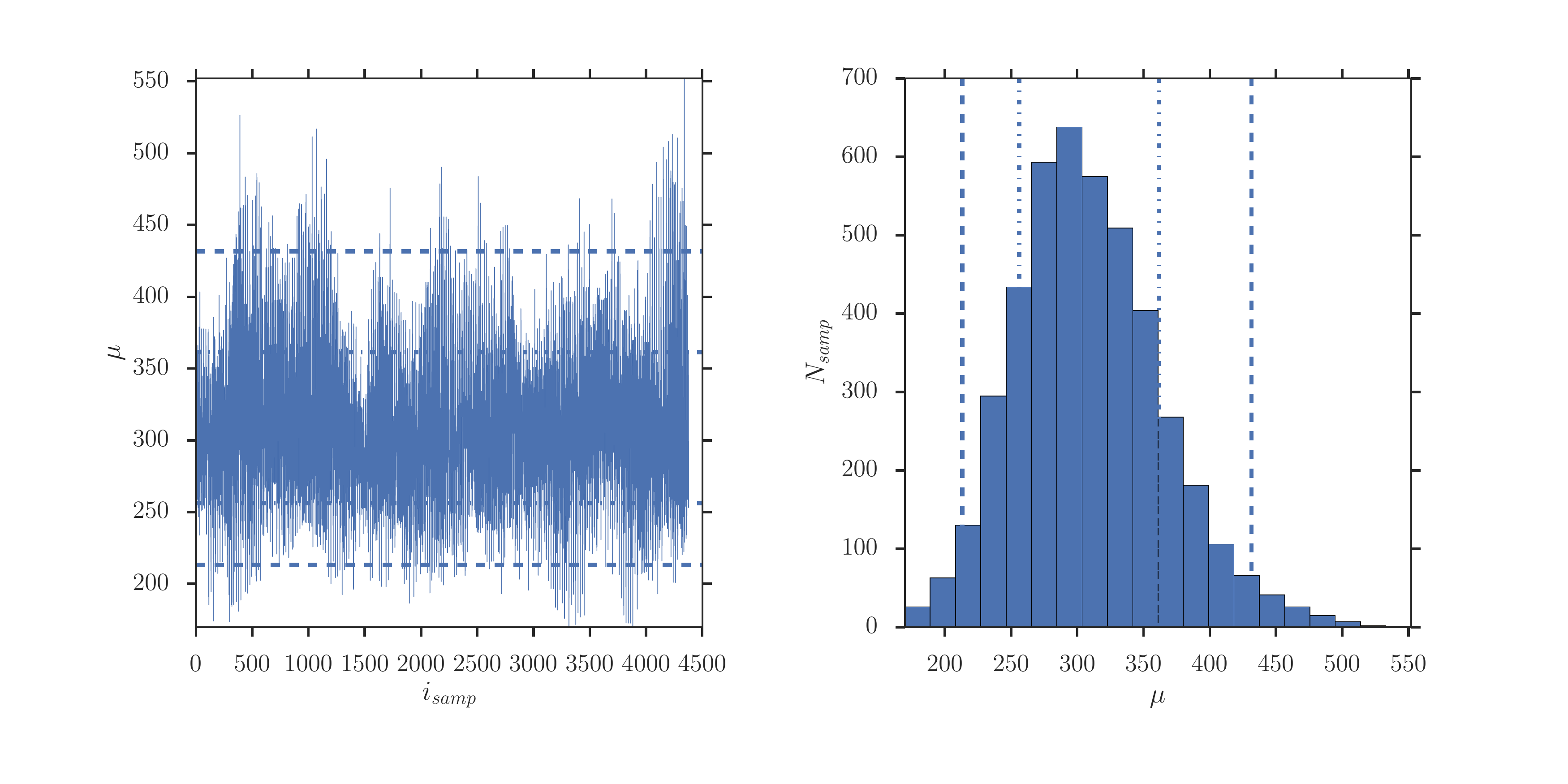}
    \caption{The posterior of the normalization of the flux distribution function obtained for the mock data run.}
    \label{figr:mockfdfnnorm}
\end{figure}

\begin{figure}[ht]
    \centering
    \includegraphics[width=0.5\textwidth]{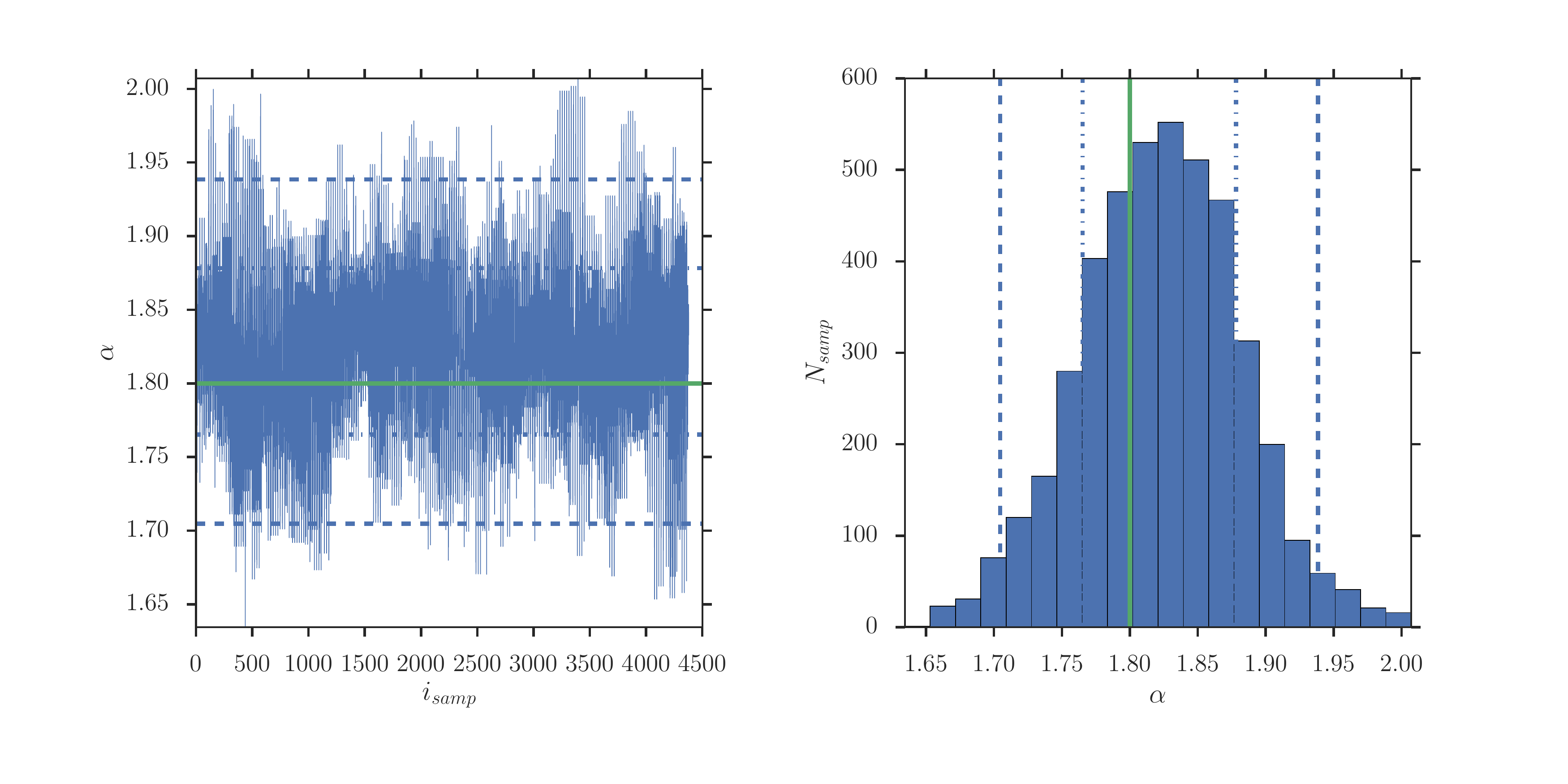}
    \caption{The posterior of the power law slope of the flux distribution function obtained for the mock data run. The green line shows the parameter value of the mock (true) model.}
    \label{figr:mockfdfnslop}
\end{figure}

We then plot the posterior distribution of the flux distribution normalization and power law slope (Figures \ref{figr:mockfdfnnorm} and \ref{figr:mockfdfnslop}). Similarly, the posterior distribution of the hyperparameters cover the true values. For the mean number point sources, the relevant true value is the imposed number of mock point sources, which is 300.

\begin{figure}[ht]
    \centering
    \includegraphics[width=0.5\textwidth]{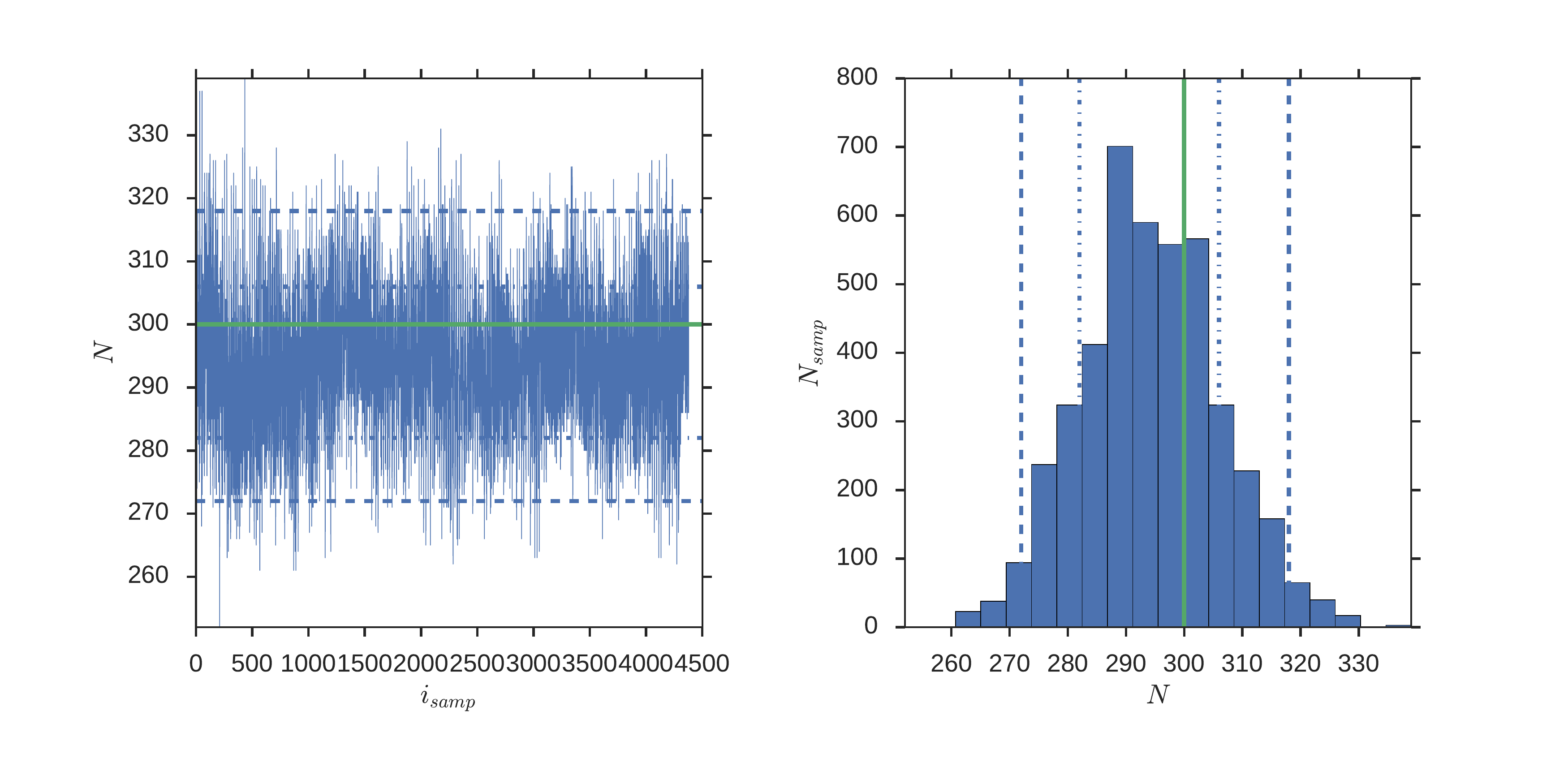}
    \caption{The posterior of the number of point sources obtained for the mock data run. The green line shows the true number of point sources.}
    \label{figr:mocknumbpnts}
\end{figure}

Lastly, Figure \ref{figr:mocknumbpnts} shows the posterior distribution of the model indicator, i.e., the number of point sources in the sample catalogs. This distribution is formally proportional to the posterior probability of the models assuming that the metamodel is true. Therefore it can be used to calculate the relative evidence (Bayes factor) for any two models.

\section{Application to Fermi-LAT Data}
\label{sect:ngal}

Next, we show results using the gamma-ray data as measured by the Fermi-LAT instrument. In order to be able to compare with the 3FGL traditional catalog published by the Fermi-LAT Collaboration, we used the reprocessed \texttt{Pass 7}, \texttt{source} class data and generated \texttt{HealPix} sky maps with base resolution 256, in three energy bins between 0.3 - 1 GeV, 1 - 3 GeV and 3 - 10 GeV. Then we constructed a joint-likelihood by fitting the front and back converted sky maps separately and taking the total likelihood. This allows a more precise modeling of the PSF, since photons that convert at the top of the instrument, i.e., front-type events, have better angular reconstruction. The choice of energy binning coincides with that of the 3FGL catalog, in order to facilitate the comparison. Furthermore, since most of the sources at high galactic latitudes are time-variable blazars, we use data in the same time interval as that used to construct the 3FGL catalog, i.e., weeks 9 through 217. 

\begin{figure}[ht]
    \centering
    \includegraphics[width=0.45\textwidth]{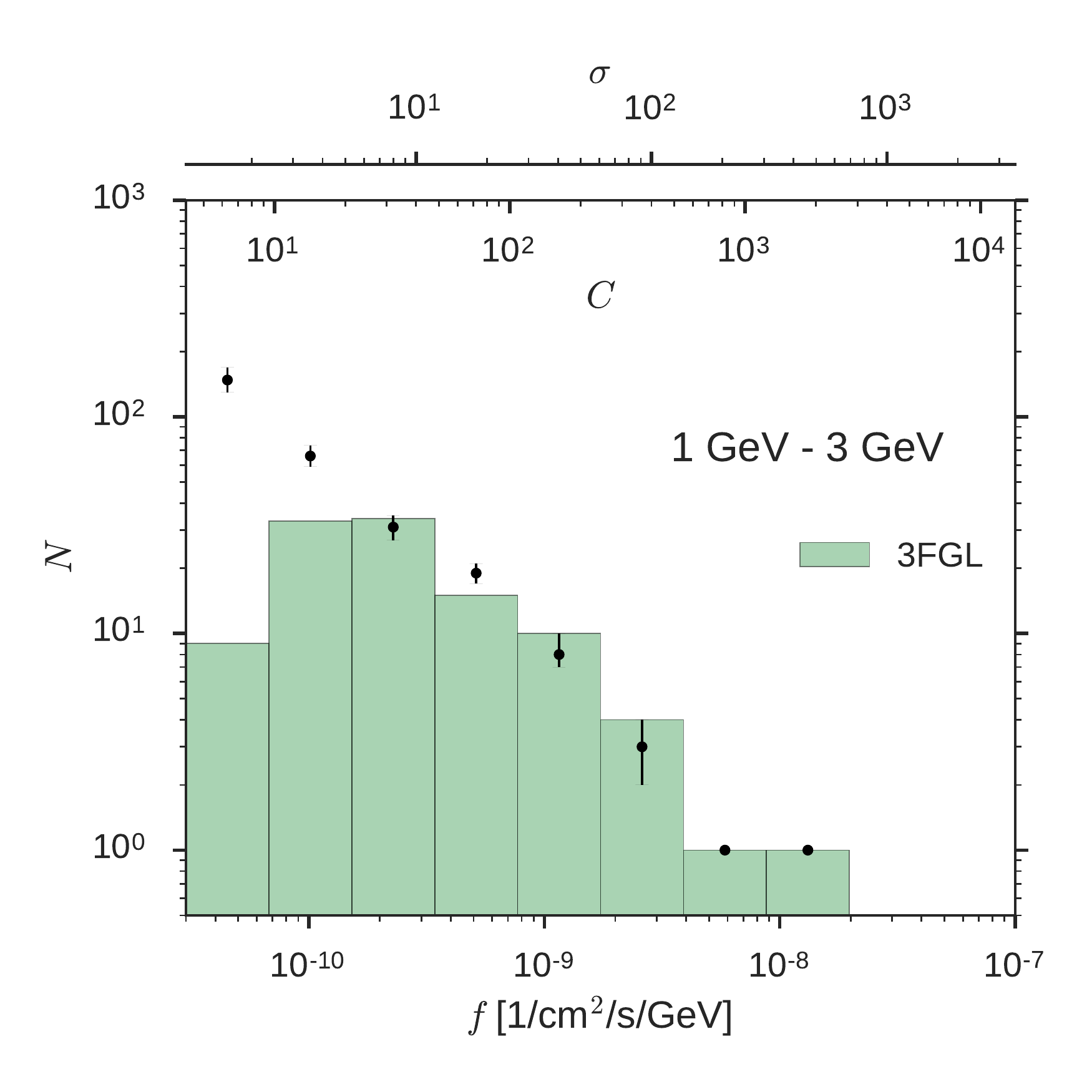}
    \caption{Posterior flux distribution function as in Figure \ref{figr:mockhistflux}. This time the green histogram shows the distribution of the 3FGL points sources.}
    \label{figr:inpthistflux}
\end{figure}

\begin{figure*}[ht]
    \centering
    \includegraphics[width=\textwidth]{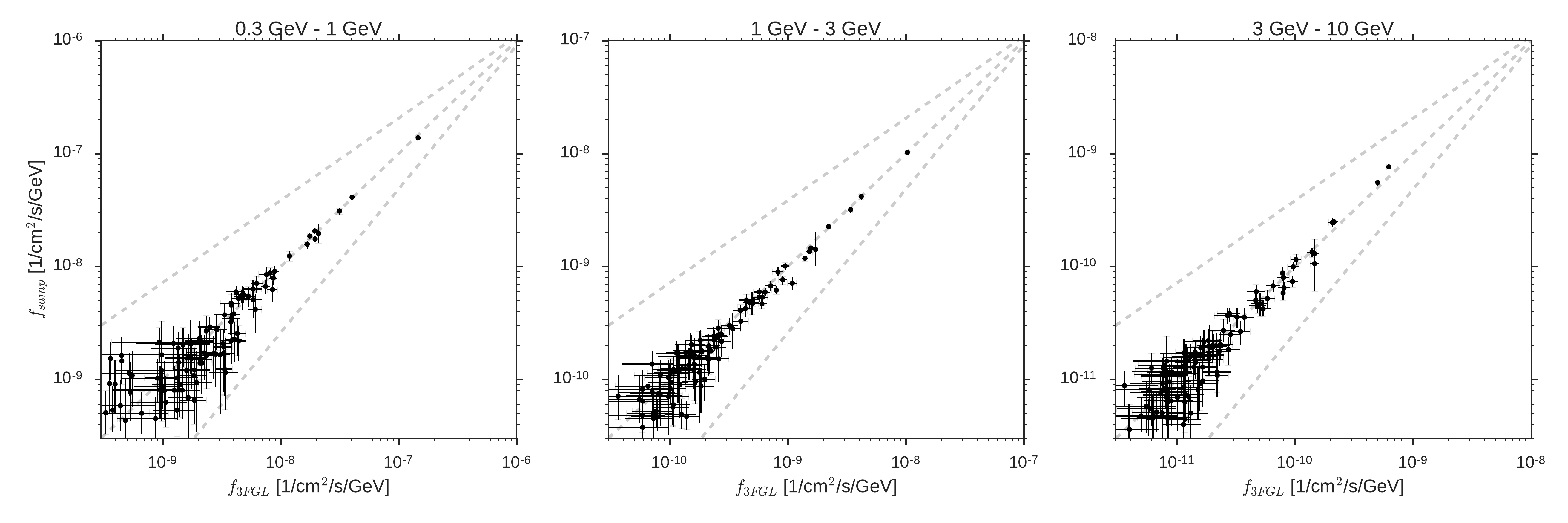}
    \caption{Association of the probabilistic catalog with the 3FGL using the same procedure as in Figure \ref{figr:mockscatspec}. Small departures from a perfect correlation with the 3FGL is partly due to different spectral modeling and partly due to the fundamentally different statistical approaches used to generate the catalogs.}
    \label{figr:inptscatspec}
\end{figure*}

A probabilistic catalog, by construction, cannot be reduced to a single list of point sources. Nevertheless its statistical summary can still be compared to a traditional catalog. This provides a means of assessing the performance of our ensemble of catalogs against the well established 3FGL. As in the mock data case, Figures \ref{figr:inpthistflux} and \ref{figr:inptscatspec} show the flux distribution function and associations between the 3FGL and probabilistic catalog.

The association between the 3FGL and the probabilistic catalog indicates that there is an agreement between the two. The correlation is stronger at larger fluxes, where covariances with the background normalization and PSF affect flux determinations negligibly. The level of agreement decreases towards lower fluxes since covariances with the background level and the radial profile of the PSF as well as the shape and normalization of the flux distribution widens the prediction of the probabilistic catalog. It is also worth noting that the scatter in Figure \ref{figr:inptscatspec} is partially due to the different spectral modeling used by the two catalogs. In other words, even if the flux predictions of the two catalogs were perfectly correlated in the pivot energy bin, 1 GeV - 3 GeV, fluxes in the other energy bins would show some dispersion.

\begin{figure}[ht]
    \centering
    \includegraphics[width=0.45\textwidth]{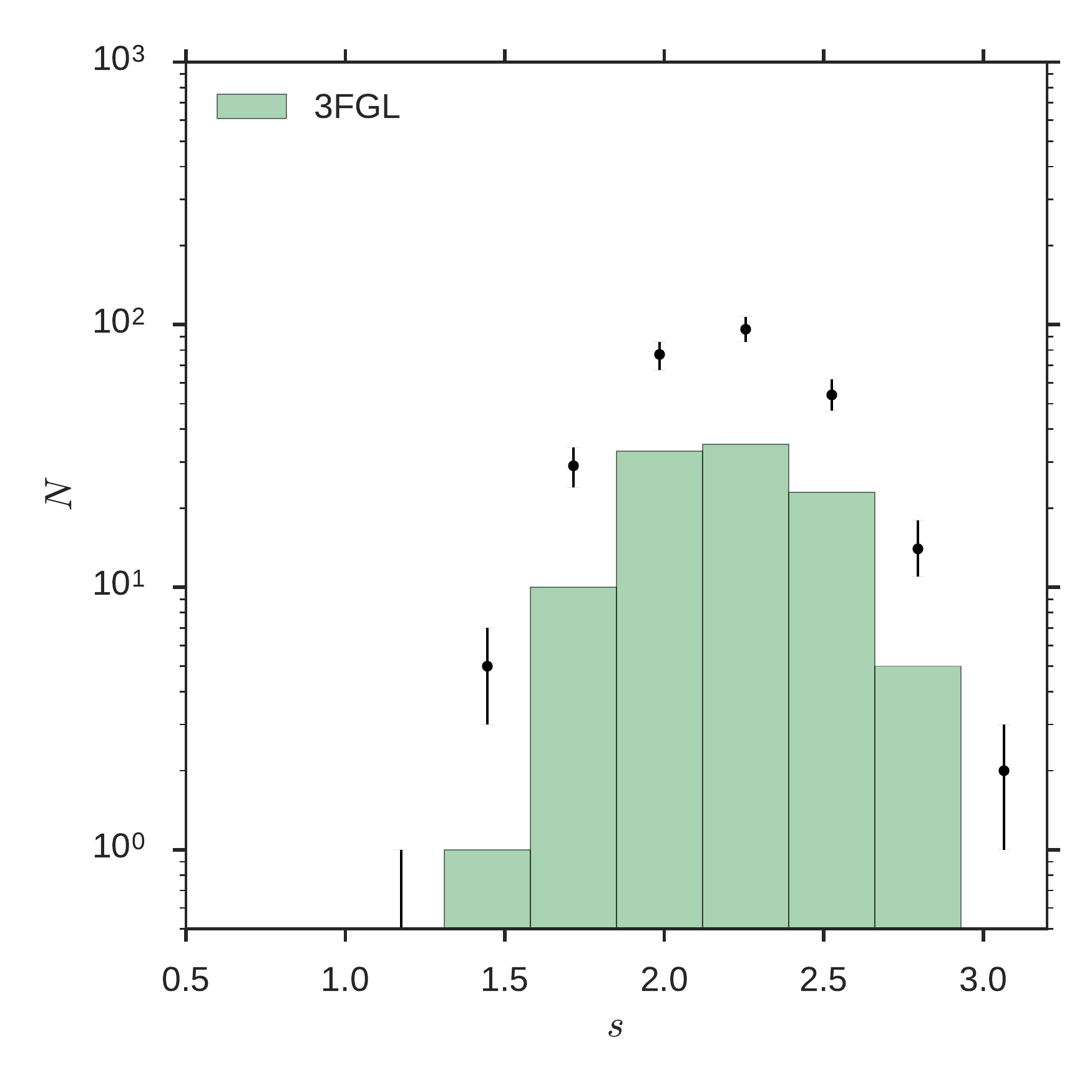}
    \caption{Posterior distribution of the point source colors.}
    \label{figr:inpthistsind}
\end{figure}

Determining the nature of a point source population, i.e., whether it is a pulsar or AGN and what subclass it belongs to, requires the reconstruction of its light curve and color distribution. We show the latter in Figure \ref{figr:inpthistsind}. The distribution around 2.2 implies that most of the point sources are blazars, where the upper tail is dominated by Flat-spectrum Radio Quasars (FSRQs) whereas BL Lacs make up the harder sub-population \citep{Dermer2016}.

\begin{figure}[ht]
    \centering
    \includegraphics[width=0.45\textwidth]{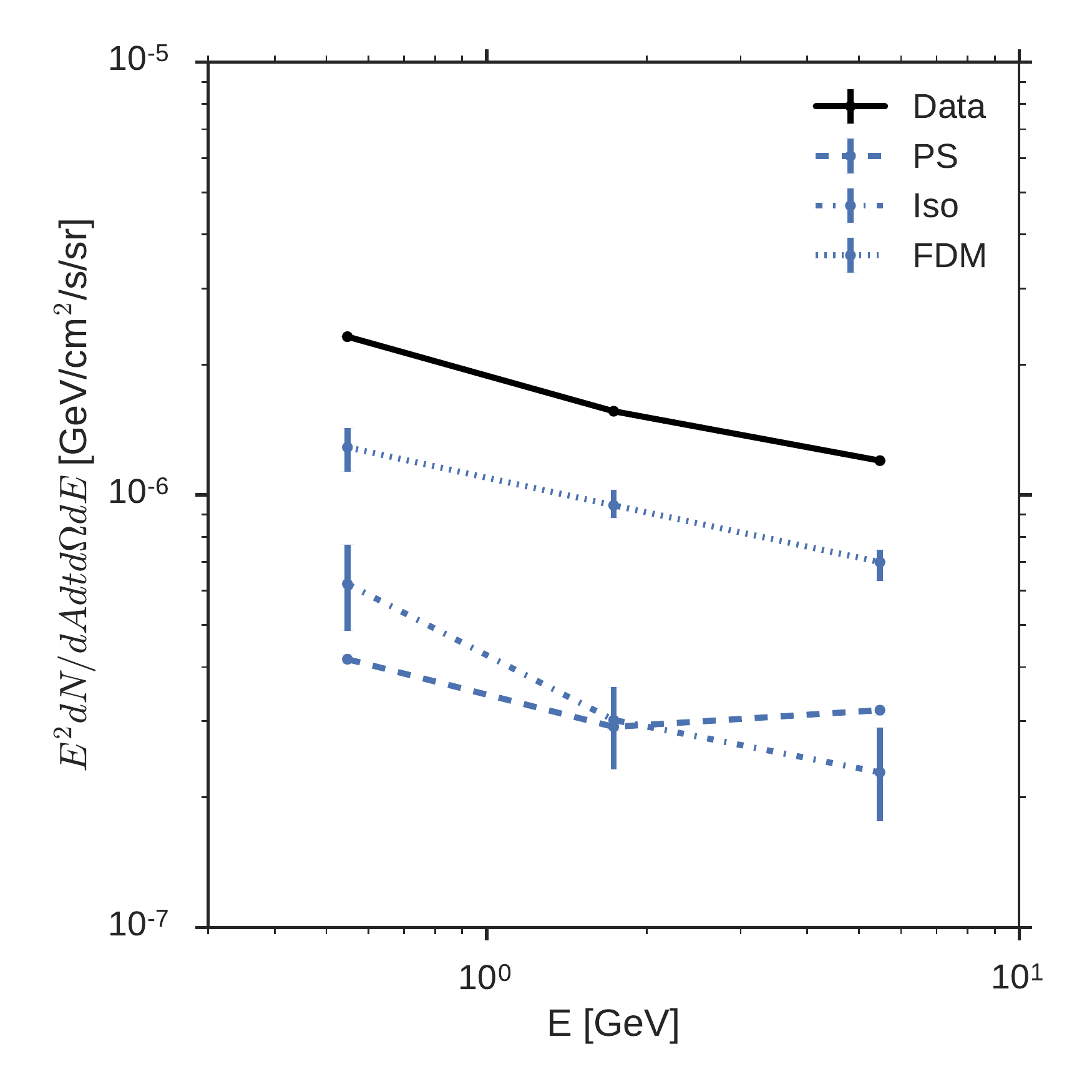}
    \caption{Spectra of emission correlated with the Fermi diffuse model (dotted), isotropic model (dot-dashed) and the total emission from point sources (dotted) averaged over the ROI.}
    \label{figr:inptcompfracspec}
\end{figure}

By collecting the point source fluxes in the ensemble of catalogs, we also estimate the contribution of the point sources to the total emission in the NGPC. The median spectra of the point sources, isotropic component, and diffuse model are given in Figure \ref{figr:inptcompfracspec}. The diffuse model is observed to be the dominant component, accounting for $57\%\substack{+6 \\ -5}$ of the total emission. The isotropic component and the point sources account for the rest in roughly equal amounts, i.e., $25\%\substack{+4 \\ -3}$ and $18\%\substack{+2 \\ -2}$, respectively. However we note that the partitioning between the point sources and the isotropic component is set by our choice of the minimum allowed point source flux. If we lower the minimum flux allowed for the point sources, then the relative contribution of the point sources would account for some of the isotropic component. The reason for this near-degeneracy is that allowing the sampler to populate the image with point sources much fainter than the level of Poisson fluctuations of the background is equivalent to decreasing the isotropic background emission. In this sense, the question of how much of the emission is accounted for by the point sources should be addressed with reference to a particular minimum point source flux.

\begin{figure}[ht]
    \centering
    \includegraphics[width=0.45\textwidth]{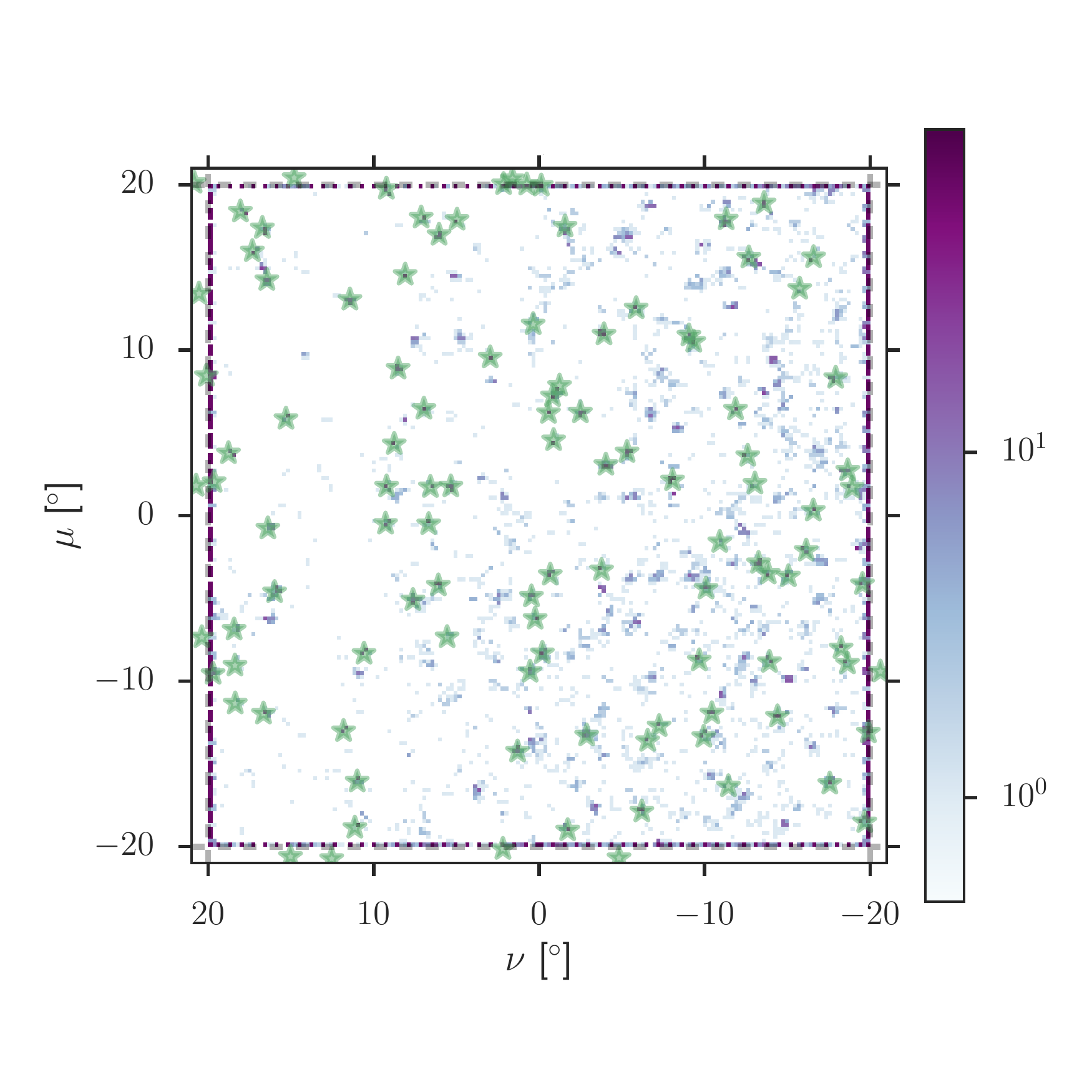}
    \caption{Catalog samples binned spatially and stacked spectrally, giving the number of point sources in fair samples from the catalog space per pixel. Green stars indicate the locations of the true point sources.}
    \label{figr:pntsbind}
\end{figure}

Samples from the catalog space can be stacked together by binning catalog samples in space and flux. This yields a map of the probability of finding a point source at a certain direction and flux, given our model. In general, this should not be interpreted as an unconditional probability, since the model used may not contain the set of all backgrounds and point sources that could be consistent with the observed data. In this work, however, the background is relatively featureless and the allowed flux distribution covers nearly four decades with a floating power law slope. This makes the stacked posterior approximately equal to the probability of finding a point source at a certain direction and flux. We show in Figure \ref{figr:pntsbind}, the catalog samples binned spatially and stacked spectrally. The color scale gives the number of catalog samples, where a model point source lands in the associated pixel. All green stars, which show the locations of the true point sources are associated with model point sources. The hot pixels away from the true point sources show regions that have $\sim1-4\sigma$ count features.

\begin{figure}[ht]
    \centering
    \includegraphics[width=0.45\textwidth]{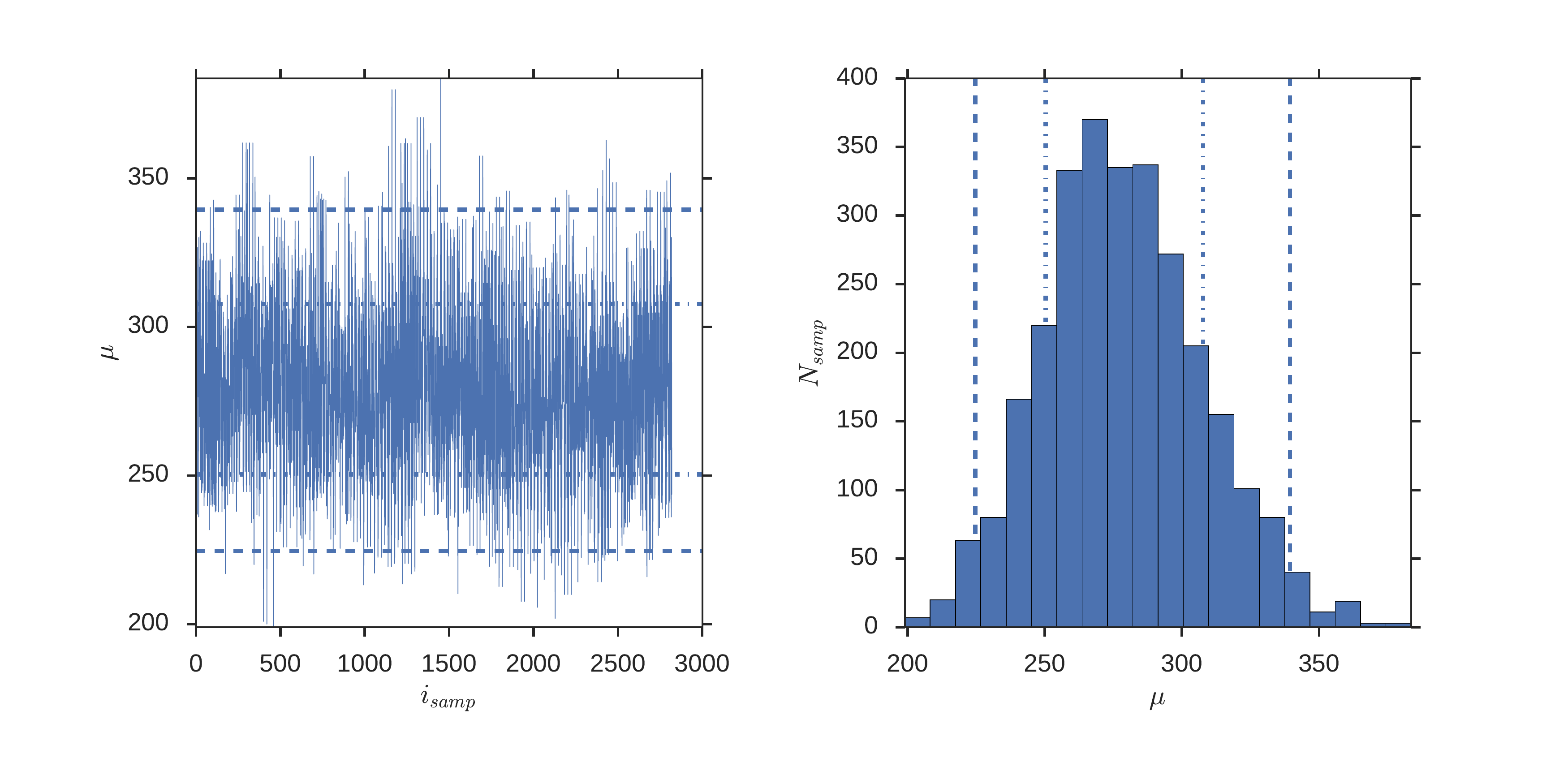}
    \caption{Posterior of the hyperprior on the normalization of the flux distribution for the real data run. The plot on the left is the MCMC time evolution of the quantity, whereas the right-hand side shows the histogram of the samples.}
    \label{figr:inptfdfnnorm}
\end{figure}

\begin{figure}[ht]
    \centering
    \includegraphics[width=0.45\textwidth]{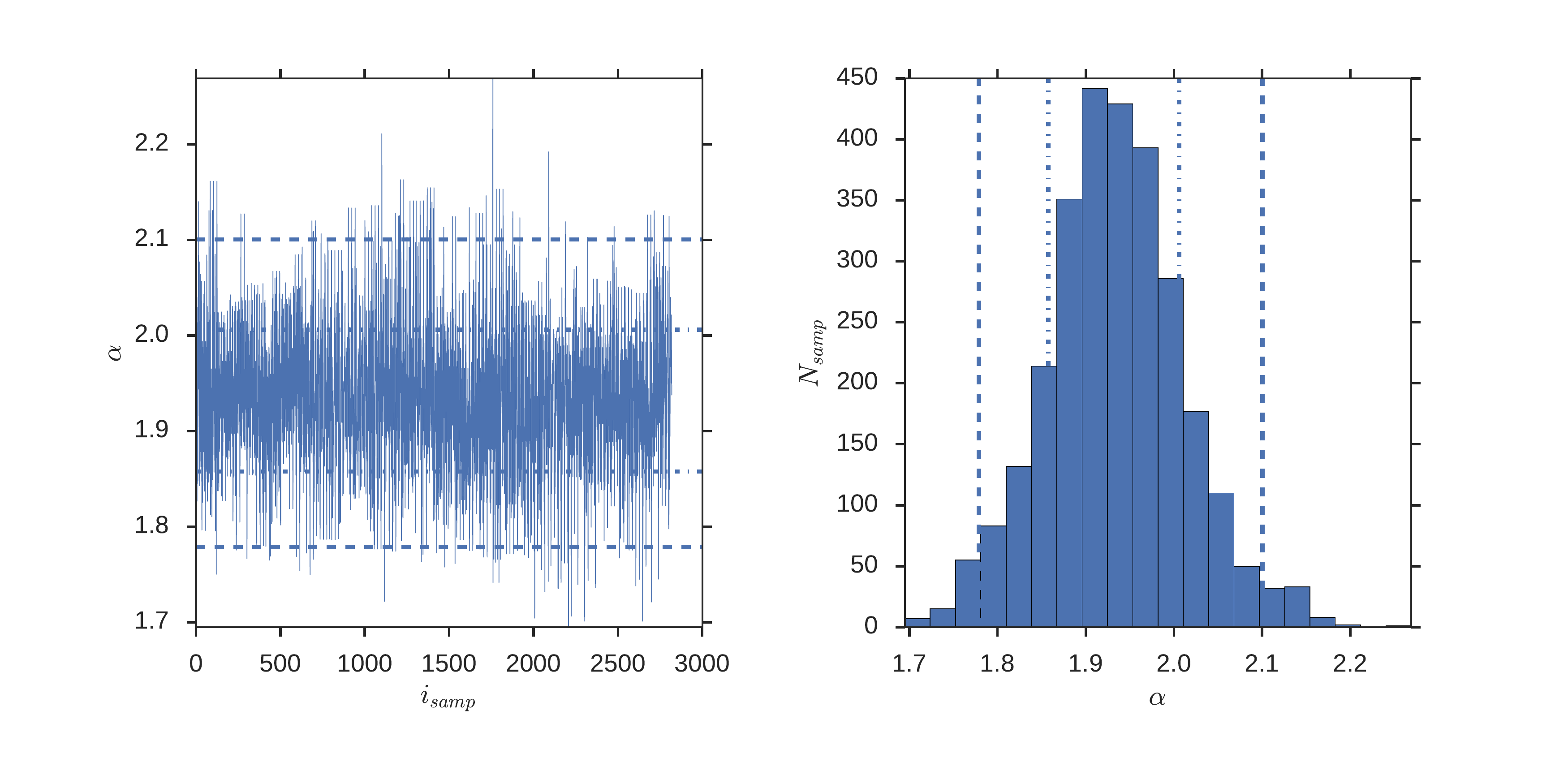}
    \caption{Posterior of the power-law slope of the flux distribution function for the real data run.}
    \label{figr:inptfdfnslop}
\end{figure}

The posterior distribution of the hyperparameters provides another handle on the population characteristics. For example Figures \ref{figr:inptfdfnnorm} and \ref{figr:inptfdfnslop} show the posterior of the normalization and power-law slope of the flux distribution function, respectively. Note that hyperparameters do not directly affect the likelihood. However their posterior distribution is still informed by the data. This is because the hyperparameter updates respect detailed balance with respect to the Poisson probability of observing the sampled model flux distribution and, in turn, model point source updates respect the Poisson probability of observing the data.

We find the power law slope to be $-1.92\substack{+0.07 \\ -0.05}$. This is smaller than the expectation from a uniform distribution of equally bright blazars, i.e., 
\begin{equation}
    \dv{N}{f} = \dv{N}{r}\Bigg(\dv{f}{r}\Bigg)^{-1} \propto r^2 \times \Bigg(\frac{1}{r^3}\Bigg)^{-1} = f^{-5/2}.
\end{equation}
Previously, \citep{Abdo2010} found that the source count function has a slope of $-2.6 \pm 0.2$ at the bright end, which hardens to $-1.6 \pm 0.1$ at the faint end. Because we use a single power law, the resulting posterior converges to an intermediate value. Moreover, given that we perform sampling only over a $1600$ degree$^2$ patch around the NGP, this analysis is subject to more shot noise. We leave a large ROI, high latitude sampling to future work.

As for the mock data, we show in Figures \ref{figr:inptnormisot} and \ref{figr:inptfdfnnorm}, the normalization of the isotropic template and the Fermi diffuse model for each energy bin, respectively. We find the median of the isotropic and diffuse backgrounds to be larger than unity by a factor of $\sim 1.3$ and $\sim 1.1$. Moreover, because the diffuse model is relatively featureless in the NGPC, the two normalizations have a large covariance.

\begin{figure}[ht]
    \centering
    \includegraphics[width=0.5\textwidth]{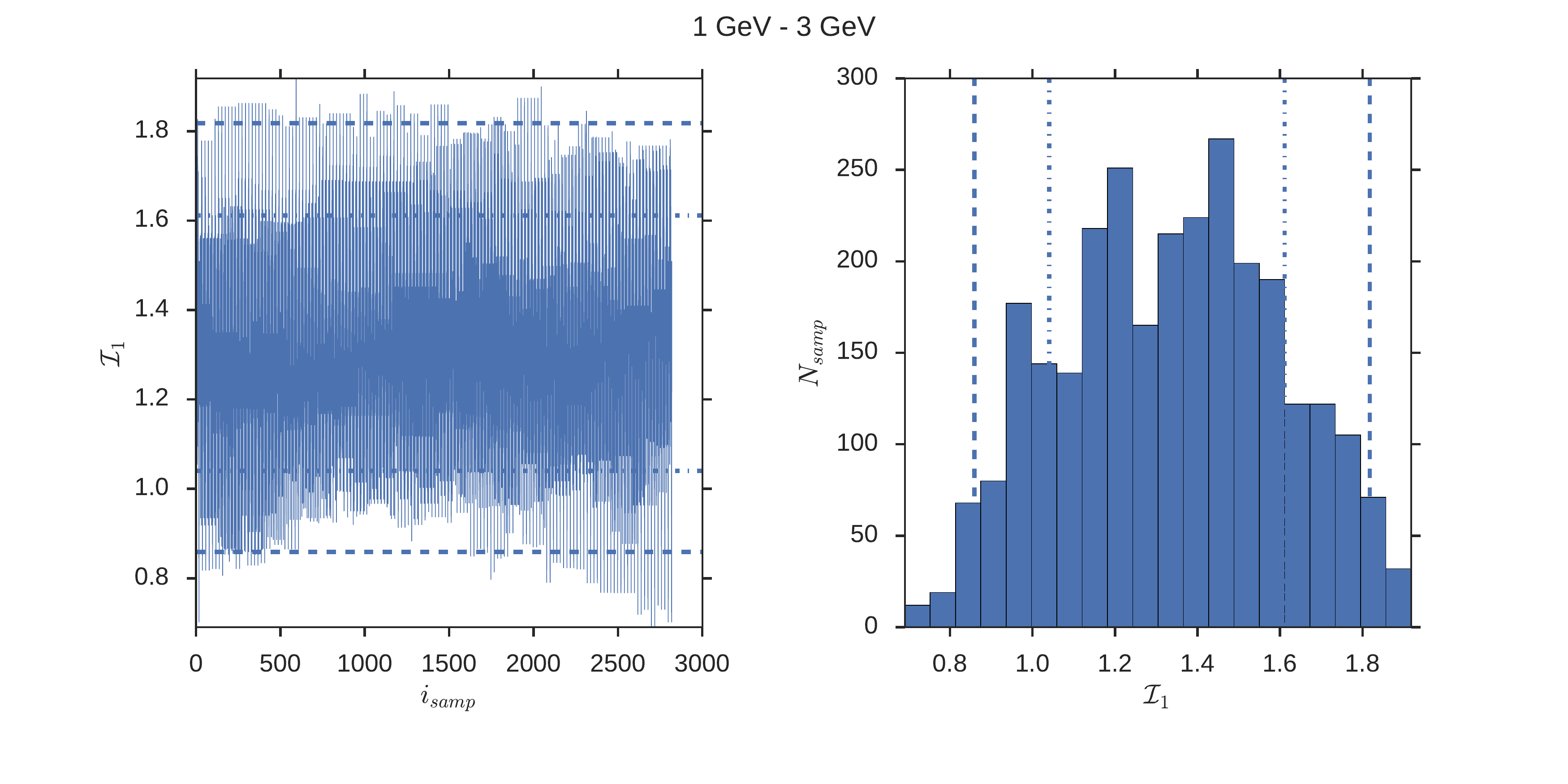}
    \caption{The posterior probability distribution of the normalization of the isotropic template obtained for the real data run.}
    \label{figr:inptnormisot}
\end{figure}

\begin{figure}[ht]
    \centering
    \includegraphics[width=0.5\textwidth]{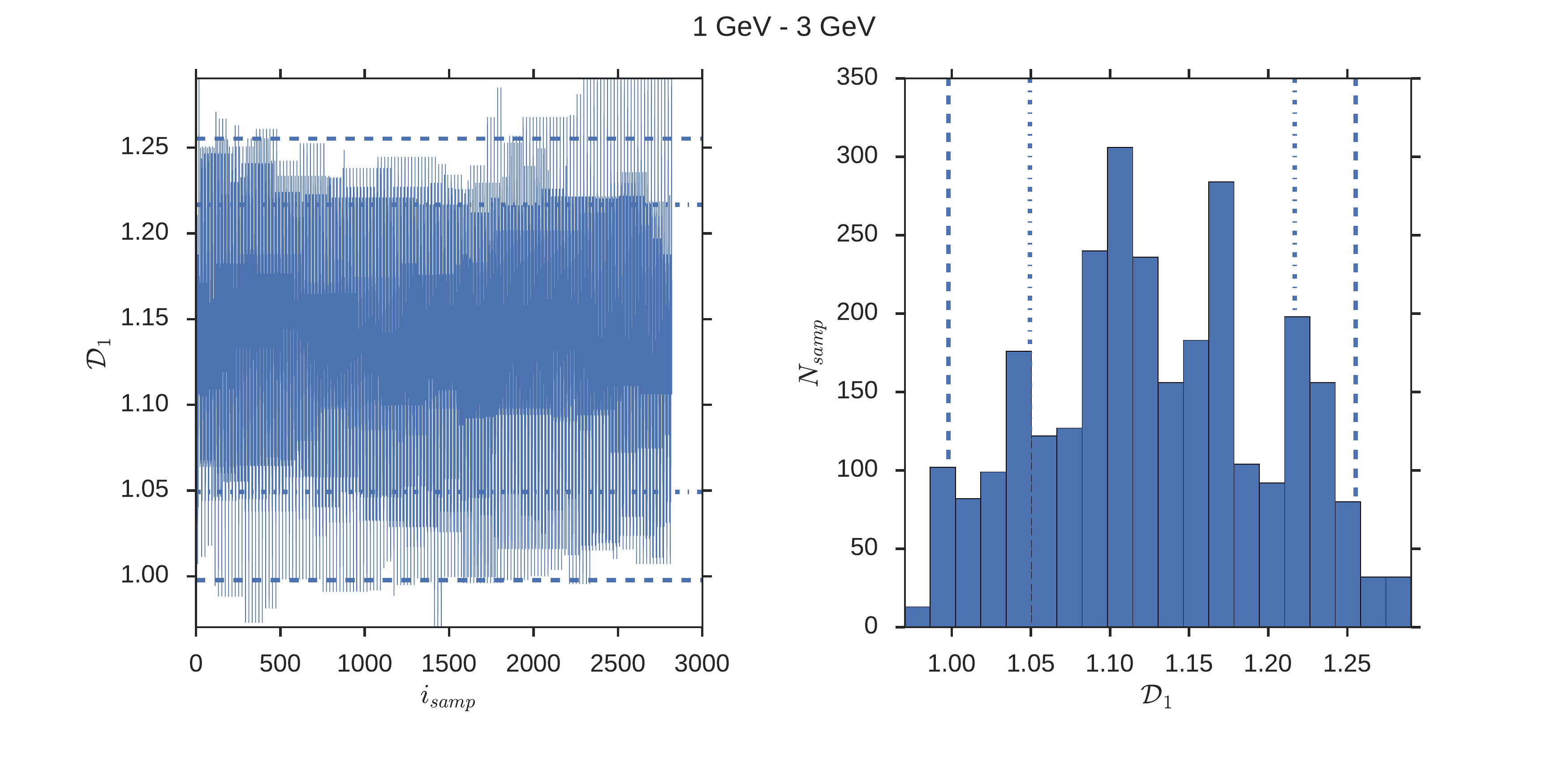}
    \caption{The posterior probability distribution of the normalization of the diffuse background template obtained for the real data run.}
    \label{figr:inptnormisot}
\end{figure}

\begin{figure}[ht]
    \centering
    \includegraphics[width=0.5\textwidth]{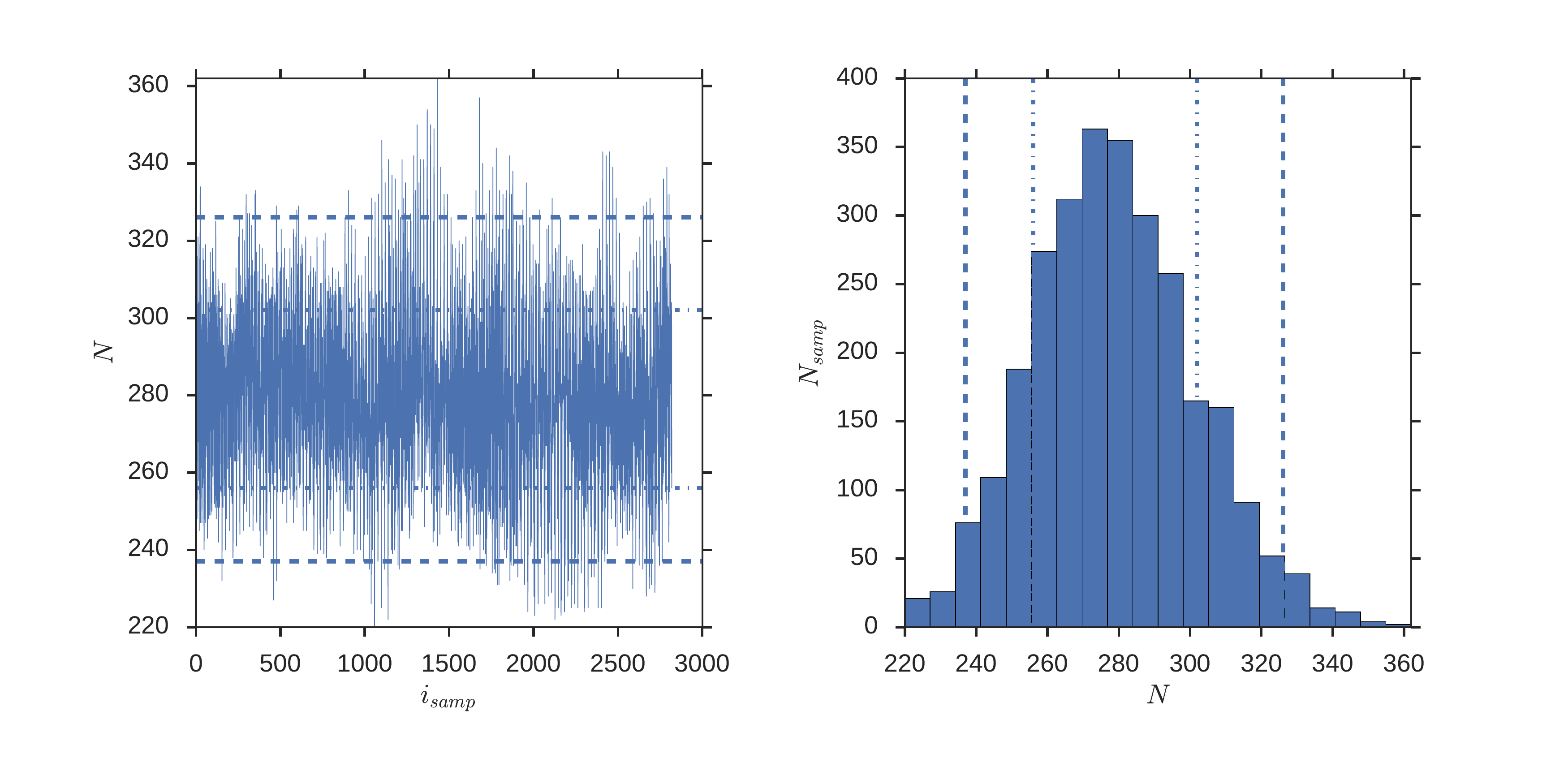}
    \caption{The posterior of the number of point sources obtained for the real data run.}
    \label{figr:inptnumbpnts}
\end{figure}

Lastly, the Figure \ref{figr:inptnumbpnts} shows the posterior distribution of the number of point sources. We infer that there are $270\substack{+30 \\ -10}$ point sources in the ROI above $3 \times 10^{-11}$/cm$^2$/s/sr/GeV at the $1-3$ GeV energy bin.

We make our probabilistic catalog available at the Harvard Dataverse \url{https://dataverse.harvard.edu} and refer the reader to Appendix \ref{sect:fits} for details on the data format.

\section{Discussion}
\label{sect:disc}

The Bayesian approach to point source inference allows a more robust treatment of the parameter covariances compared to finding the most likely catalog. Furthermore, information encoded in the subthreshold features in the image is not discarded. An important motivational distinction between conventional and probabilistic cataloging is that the latter aims to find a set of point sources that is free of false positives, whereas member point sources in the former are not guaranteed to exist. It is the repeated sampling of the raw data along with the false positives that make probabilistic cataloging a very useful tool in probing population characteristics.

The median flux of spatial associations of our catalog with the 3FGL agrees remarkably well when associated with the 3FGL fluxes. This demonstrates the feasibility and reliability of our probabilistic approach. We further verify the performance of probabilistic sampling in the unresolved regime by using mock data and demonstrate that the truth information is covered by our posterior.

With the current implementation several thousand realizations of even full sky gamma-ray catalogs such as the 3FGL or 3LAC \citep{Ackermann2015}, which is an AGN catalog based on the 3FGL, can be generated. In fact, given the potential increase of computational resources available to catalog generation in astrophysics, it may even be feasible to make the next generation standard catalog reduction pipelines, probabilistic. As for the time domain analysis, FAVA \citep{Ackermann2013}, does not use a likelihood approach and therefore cannot be generalized using probabilistic cataloging. Given the time complexity of catalog sampling, it is unlikely that full sky catalogs can be sampled by binning data in time domain. Furthermore, flares or periodic flux variations of individual point sources will likely be washed out in the flux distribution. Since probabilistic sampling is most informative when studying the population characteristics, it is not particularly useful for time domain analyses. It may also be possible to generalize our method to other wavelengths such as optical datasets. However typical PSF size of less than an arcsecond in optical photometry restricts the application of probabilistic cataloging to $\sim$ deg$^2$ sized ROIs due to the large number of pixels. See \citep{Portillo2017} for an example of this technique applied to optical photometry in a crowded stellar field.

When multiple samples are drawn from the catalog space, point sources in one sample catalog cannot simply be matched to those in other samples. This is due to the fact that the underlying likelihood function is invariant to permuting the labels of model point sources. This is known as the labeling degeneracy problem \citep{Zhu2014} and is common to all mixture models, where model components are not individually labeled. Therefore taking the ergodic average of a given model parameter without breaking this degeneracy becomes meaningless, since parameters change identity during the evolution of the chain. For instance in the limit of an infinitely long chain, the ergodic average of all flux parameters would be identical and, hence, not useful for inferring the flux posteriors of individual point sources. Nevertheless, inference of population characteristics does not necessitate breaking of the labeling degeneracy, since population characteristics are also invariant under permutations of point source labels. The time we need to explicitly break the labeling degeneracy is when comparing the chain of catalog samples to a traditional catalog. This requires a prescription to associate point sources. Associations at low fluxes inevitably causes loss of information and is the major reason for the increasing spread in Figures \ref{figr:mockscatspec} and \ref{figr:inptscatspec} towards low fluxes.

Model choice in the Bayesian framework tries to establish a balance between the goodness of fit of a model and its predictive power. A point source model will fit a given count map at least as well as another with fewer point sources. In fact, in the limit of very large number of point sources, the point source model essentially becomes indistinguishable from a diffuse model with arbitrary number of spatial degrees of freedom. It is thus desirable that the chosen model covers the smallest possible volume when projected onto the data space while still fitting the observed data reasonably well. Known as the Occam's razor, this principle is encoded in the marginal likelihood. This quantity penalizes point source models for wasted parameter space, i.e., for only marginally increasing the goodness of fit at the expense of predicting extraordinary data, which significantly reduces the likelihood for most of the added parameter space.

The difficulty in probing faint point sources is the inability of the Poisson-noise count data in constraining point sources fainter than the typical Poisson fluctuations of the background. This sets a fundamental count scale, $\sqrt{C_B}$, where $C_B$ is the mean number of counts expected from the background inside a FWHM for the given exposure. This scale necessitates a careful choice of the minimum point source flux allowed by the model, $f_{min}$. Let us denote the number counts that corresponds to $f_{min}$ by $C_{min}$ for the given exposure. In the limit $C_{min} \gtrsim \sqrt{C_{B}}$, the goodness of fit of the point source model gets reduced due to lack of favorable parameter space. On the other hand, if $C_{min} \lesssim \sqrt{C_{B}}$, then the data cannot constrain the parameter space. In order to assess the regime of a given prior choice, we evaluate the relative information gain in going from prior to the posterior and choose $f_{min}$ such that the posterior is sufficiently informed by the data. See Appendix \ref{sect:info} for a more detailed discussion.

A well known problem of Poisson regression of photon count data is that of mismodeling. It is important to note that when a test statistic of maximum log-likelihood difference between the alternative and null models is interpreted as a detection significance, it is implicitly assumed that the underlying model is a good description of the data. Otherwise the unquoted systematic uncertainty can be much larger than the statistical uncertainty. For example, when a flux-incomplete point source model is used, this assumption will not hold true. Therefore predictions for the diffuse templates will be biased so as to minimize the residuals of the flux-incomplete point source model. Probabilistic cataloging addresses this problem by shunting uncertainties due to flux-incompleteness to a pure statistical form. This is accomplished through sampling in the catalog space above a given cutoff flux, which contains all point source configurations that the data is consistent with. By doing so, it allows many false-positives in the model. But these are sampled less frequently compared to the true-positives as long as the sampling is performed in the likelihood-dominated region. Note that probabilistic cataloging can still suffer from mismodeling due to imperfect background models.

In this work, we follow Bayesian statistics to perform inference with the motivation that models can be penalized for introducing unnecessary point sources. Strictly speaking a frequentist approach that adds a degree of freedom penalizing term to the test statistic in the form of Akaike Information Criterion (AIC) \citep{Akaike1974} or Bayesian Information Criterion (BIC) \citep{Schwarz1978} could achieve the same goal by iterating the fit over a range of number of point sources. However in the limit of faint point sources, the likelihood topology becomes nearly degenerate and the maximum likelihood becomes a poor indicator of the goodness of fit of a model. 

In \citep{Brewer2012} a similar sampler was implemented and shown to work on mock data using a single energy bin. The sampling was performed directly on the prior with a hard likelihood threshold that depends on a series of levels whose prior mass decreases successively. Known as the Diffusive Nested Sampling (DNS) \citep{Brewer2009}, this method can sample multimodal distributions with high likelihood contrast. Furthermore, as a by-product, it provides an estimator for the Bayesian evidence. In this work, however, we sample from the posterior, since we do not need the Bayesian evidence for the point source metamodel. Instead, the model choice is based on the relative frequency of visiting different models via transdimensional jumps. In addition, the likelihood topology of the problem, although being highly multi-modal, has shallow islands since most point sources are faint. This argument excludes the well-localized bright point sources. Efficient exploration of these bright members would be impossible by Metropolis-Hastings updates alone. However we employ splits and merges for this reason in order to facilitate chain mixing in the crowded field limit. Similarly, \texttt{HELP: XID+} \citep{Hurley2017} is another Bayesian point source extraction framework that can fold in prior information on the sources. Although both algorithms are tailored for the crowded limit with a rich covariance structure, \texttt{HELP: XID+} differs from \texttt{PCAT} in that it relies on the prior knowledge of a certain number of known source positions in order to infer flux and flux uncertainties of sources consistent with the observed data.

\section{Conclusion}
\label{sect:conc}

In this paper, we implemented a transdimensional sampler to infer point sources in a given photon count map by sampling from the probability distribution of the underlying catalog space. Our approach allows a consistent Bayesian exploration of the background normalization and the PSF, where they are co-sampled along with the model point source positions, fluxes and spectral parameters. The output is an ensemble of catalogs consistent with the count data, which represents the probability distribution of the catalog given the image.

Compared to the traditional maximum likelihood solution to point source inference, probabilistic cataloging is computationally expensive. However this is a price paid for superior control over covariances in the catalog space, which becomes critical when a subsequent analysis using the inferred catalog as an input tries to reach a conclusion. Instead of using the maximum likelihood catalog with Gaussian errors around this solution, probabilistic cataloging provides a robust treatment of uncertainties.

As a case study we apply our technique first to mock data and then to gamma-ray data in the NGPC. We associate the resulting probabilistic catalog with the 3FGL, and measure the flux distribution function down to $\sim$ 1-sigma sources. 

\acknowledgments

We thank David W. Hogg, Brendon Brewer, Benjamin Lee, Zachary Slepian, Greg Green, Albert Lee, Can G\"{o}kler and Emre Erge\c{c}en for useful discussions during the course of the project.

\newpage

\appendix
\section{RJMCMC proposals}
\label{sect:accpprob}
In this Appendix we expand on the implementation details of various proposal types and their acceptance ratio calculation as given in \ref{equa:accp}.

In order to explore the catalog space we make a set of proposals given in Table \ref{tabl:proptype}. The name of the proposal is given along with its weight, which is proportional to its frequency, where $N_{max}$, $N_{psf}$ and $N_e$ are the maximum number of point sources, number of PSF classes and number of energy bins, respectively. The motivation behind this choice is to share the proposals equally between each parameter in the parameter vector.

\begin{table}[ht]
    \caption{Types of proposals used to explore the catalog space.}
    \centering
    \begin{tabular}{c c c c}
        Name & Weight \\
        \hline\hline
        Point source parameter updates & $4N_{max}$ \\
        \hline
        PSF parameter updates & $4N_{psf}N_{e}$ \\
        \hline
        Background normalization updates & $2N_e$ \\
        \hline
        Birth and death & $N_{max}$ \\ 
        \hline
        Split and merge & $0.2N_{max}$ \\ 
        \hline
        Hyperparameter updates & 2 \\
        \hline
    \end{tabular}
    \label{tabl:proptype}
\end{table}

\paragraph{Point source parameter updates}

The point source parameter updates, which involve changing the position, flux or spectral parameters of the point sources, are the usual within-model proposals that explore the posterior probability distribution of a point source model of a given dimensionality, $N$. Even though within-model moves do not require the RJMCMC formalism, we also make use of auxiliary variables for these updates. If a parameter $\tilde{\theta}$ is to be updated using an auxiliary variable, $u$, we draw $u$ from a heavy-tailed Gaussian distribution with mean zero.
\begin{gather}
    (\tilde{\theta}, u) \to (\tilde{\theta}^\prime, u^\prime) \\
    \tilde{\theta}^\prime = \tilde{\theta} + u \\
    u^\prime = -u
\end{gather}
Here, we denote CDF-transformed variables, i.e., having a uniform distribution between 0 and 1, with a tilde. Unlike transdimensional proposals such as birth and death, within-model proposals are self-antagonist in the sense that both the forward and the reverse transition between two points in the state space can be achieved using the same type of proposal. Therefore, since there are as many parameter updates possible as the number of parameters, which is fixed during a parameter update, the combinatorial ratio in the acceptance ratio is unity. Similarly the Jacobian is unity, since the proposed state is the sum of the current state and a random offset.

\paragraph{PSF and background normalization updates}

PSF and background normalization updates are other types of within-model proposals that allow the exploration of likely covariances between the catalog space, the background normalizations and the PSF of the instrument used to collect the count data.

\paragraph{Birth and Death}

Birth and death proposals, where a point source is either added to or deleted from the point source list, are the elementary across-model moves that allow the exploration of point source models of different dimensionality. During a birth, the sampler proposes the transformation
\begin{align}
    (\theta, u) \to (\theta, \theta_*) = \theta^\prime
\end{align}
where the tuples $(\theta, u)$ and $(\theta, \theta_*)$ represent the current and proposed state of the sampler, respectively. Here, the auxiliary variable vector, $u$, coincides with the parameters of the added point source, $\theta_*$, e.g., its coordinates, flux and spectral parameters. Conversely, during a death proposal we require
\begin{align}
    \theta = (\theta^\prime, \theta_*) \to (\theta^\prime, u^\prime)
\end{align}
where the auxiliary variable $u$ carries the parameters of the point source to be killed, $\theta_*$. The relative frequency of proposing a birth move can be adjusted at the expense of inversely scaling the acceptance ratio. We do not break this symmetry and make antagonist proposals equally likely. Since the auxiliary variable vector is identical to the added point source parameters, the Jacobian also becomes unity. Note that we sample the new point source parameters from the prior, which cancels the auxiliary vector density with the prior density during both birth and death, i.e.,
\begin{equation}
    \frac{P(\theta_*)}{g(u)} = \frac{g(u^\prime)}{P(\theta_*)} = 1
\end{equation} 

When adding or deleting point sources to and from the point source list, we treat it as an ordered list. If the list of point sources was treated as an unordered list, then the number of possible deaths would equal the number of point sources, whereas the number of possible births would be one. Therefore, assuming that there are $N$ point sources at a given state, the combinatorial ratio would become $1/(N+1)$ for births and $N$ for deaths. However, in that case there would be an implicit $N+1$ fold degeneracy in the density of states, $g(u)$, in the case of birth proposals. Similarly $g(u^\prime)$ would be denser by a factor of $N$ in the case of death proposals, which would make the overall combinatorial ratio equal to unity. Therefore, a simpler picture is to treat the point source list as an ordered set, where the ratio is unity. Moreover, if the list of point sources had been treated as unordered, an issue arises since the death of a given point source could be taken as the reverse move of the birth of any point source. This causes the transformation to lose its one-to-one property, conflicting the RJMCMC condition that transformations should be bijections. Adoption of an ordered point source list alleviates this problem by making the bijective nature of the transformation manifest.

\paragraph{Splits and Merges}

Birth and death proposals can, in principle, explore the catalog space. However the exploration of parameter covariances in the crowded limit can be slow when using only such elementary proposals. We therefore require dedicated proposals types, splits and merges, in order to efficiently explore whether features in the count map are more consistent with single and bright or multiple and faint point sources.

During a split proposal a point source gets split into two point sources through the use of an auxiliary vector, $\vec{u}$, such that
\begin{align}
    \theta, u = (\theta_0, \theta_*, u) \to (\theta_0, \theta_1, \theta_2)
\end{align}
where $\theta_0$ is the part of the current state vector that does not get updated. Let the point source with parameters $\theta_*$ have the galactic longitude $l_*$, galactic latitude $b_*$, flux $f_*$ and spectral index $s_*$. Then it gets split into two point sources with fluxes $f_1$ and $f_2$ such that
\begin{align}
    l_1 &= l_* + \Bigg(1 - \frac{u_f}{f_*}\Bigg) u_l \\
    l_2 &= l_* - u_l \frac{u_f}{f_*}  \\
    b_1 &= b_* + \Bigg(1 - \frac{u_f}{f_*}\Bigg) u_b \\
    b_2 &= b_* - u_b \frac{u_f}{f_*} \\
    f_1 &= u_f \\
    f_2 &= f_* - u_f \\
    s_1 &= s_* + u_s \\
    s_2 &= s_* - u_s
\end{align}
where $u_l$, $u_b$, $u_f$ and $u_s$ are the auxiliary variables. In order for splits and merges to have high enough acceptance ratio, we require $u_l$ and $u_b$ to have compact support and be drawn from a uniform distribution between $U_{lb}$ and $-U_{lb}$. Given that the PSF of the Fermi-LAT is below 1 degree in all the energy bands of interest, we chose $U_{lb}=1^\circ$ in order to optimize the exploration of the overlapping point sources. Furthermore we draw $u_f$ and $u_s$ from the prior distribution of the flux and spectral index, respectively. 

\paragraph{Hyperparameter updates}

Hyperparameter updates are within-model moves that do not involve a likelihood change. They reparametrize the prior such that the prior on the current point source parameters gets updated. In the current implementation of \texttt{PCAT}, the only prior that admits a set of hyperparameters is that on the flux distribution, which is parametrized by its normalization and power-law slope. These hyperparameters are updated just like any other parameter using heavy tailed Gaussian proposals.

\section{Choosing $f_{min}$}
\label{sect:info}

Hierarchical modeling allows a well-motivated parametrization of the prior distribution of the point source fluxes. However if the prior region is made larger in a direction, where data is not constraining, the posterior transitions from a likelihood dominated regime to the prior dominated regime. In other words, arbitrarily faint point sources can comfortably fit a given count map while uninformed by the data. As a result, choice of priors becomes critical. In order to probe the region where this transition occurs, we compute the relative entropy of the posterior compared to the prior \citep{Jaynes1955}. Denoting the posterior and prior densities with $P(\theta|D)\equiv\dd{P}(\theta|D)/\dd{\theta}$ and $P(\theta)\equiv\dd{P}(\theta)/\dd{\theta}$, where $\theta$ is the parameter vector and $D$ is the observed data, the quantity,
\begin{equation}
\label{equa:info}
    D_{KL} = \int \dd{\theta} P(\theta|D) \ln \frac{P(\theta|D)}{P(\theta)}, 
\end{equation}
is known as the Kullback-Leibler divergence \citep{Kullback1951}. It encapsulates the amount of information lost when using the prior instead of the posterior, expressed in the natural unit of information, i.e., nat. Using the Bayes rule \citep{Bayes1958a},
\begin{equation}
    P(\theta|D) = \frac{P(D|\theta)P(\theta)}{P(D)},
\end{equation}
KL divergence can be written as an ensemble average over the realized Markov chain,
\begin{align}
    D_{KL} &= -\ln P(D) + \int \dd{\theta} P(\theta|D) \ln P(D|\theta) \\
    &= \Bigg\langle \ln P(D|\theta) \Bigg\rangle - \ln P(D).
\end{align}
The Bayesian evidence, $P(D)$, encodes the probability of obtaining the observed data under the point source metamodel. We calculate $D_{KL}$ for various values of $f_{min}$ using the same dataset.

\begin{figure}[ht]
    \centering
    \includegraphics[width=0.7\textwidth]{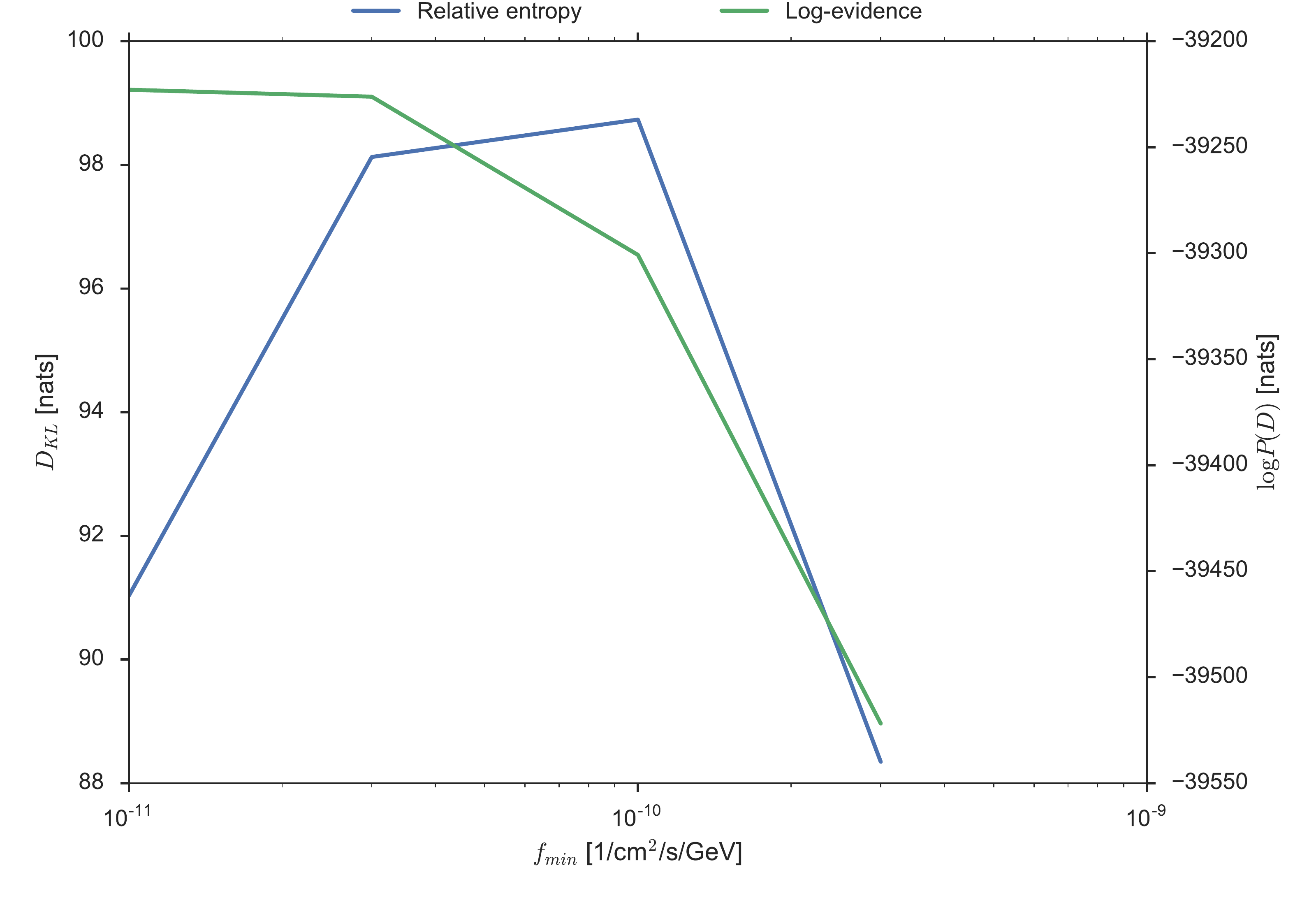}
    \caption{Information gain, Equation \ref{equa:info}, evaluated over independent posterior chains with different $f_{min}$.}
    \label{figr:info}
\end{figure}

Ideally, the choice of prior must be fixed before looking at the data. Therefore, we choose the flux prior based on the scale invariance argument and only use the KL divergence to probe when the inference problem enters the prior dominated region. In principle the expected value of the KL divergence over the set of realizations of the data could be used to determine \emph{the} reference prior \citep{Berger2009}. The $f_{min}$ that maximizes the relative information under this expectation,
\begin{align}
    \hat{f}_{min} = \argmax \avg{D_{KL}}
\end{align}
would yield the objective prior that is invariant under reparametrizations of $\theta$ as in the case of Jeffrey's prior,
\begin{align}
    D_{KL} &= \int \dd{\theta} P(\theta^\prime|D)\dv{\theta^\prime}{\theta} \ln \frac{P(\theta^\prime|D) \dd{\theta^\prime/\dd{\theta}}}{P(\theta^\prime)\dd{\theta^\prime/\dd{\theta}}} = \int \dd{\theta^\prime} P(\theta^\prime|D) \ln \frac{P(\theta^\prime|D)}{P(\theta^\prime)} 
\end{align}
but also remains uninformative in the multivariate case. However this is a computationally intractable task that requires many \texttt{PCAT} runs over simulated data. 

In order to compute the Bayesian evidence, $P(D)$, we use a truncated harmonic mean estimator
\begin{align}
    \frac{1}{P(D)} &= \int \dd{\theta} \frac{Q(\theta)}{P(D)} = \int \dd{\theta} \frac{Q(\theta)P(\theta|D)}{P(D|\theta)P(\theta)} \\
    &= \Bigg\langle \frac{Q(\theta)}{P(\theta)} \frac{1}{P(D|\theta)} \Bigg\rangle
\end{align}
where $Q(\theta)\equiv\dd{Q}(\theta)/\dd{\theta}$ is a probability density defined over parameter space with appropriate normalization. In the standard implementation, rare samples with lower likelihood have have higher weights. Therefore as more samples are taken during the evolution of the MCMC, the lowest likelihood samples, which are subject to shot noise, destabilize the estimation. By forcing $Q(\theta)$ to have compact support over a well-sampled region, we ensure that the harmonic mean estimator is free of noise. The fact that $Q(\theta)$ has steeper tails than $P(\theta)$ compensates for the truncated ensemble of samples by down-normalizing the resulting evidence.

In the limit of large $f_{min}$, the relative information gain in the posterior with respect to the prior is small, since the parameter space available to the fit is constrained. Similarly the evidence is small, since most of the observed faint features cannot be fitted well with bright point sources. However, as $f_{min}$ is decreased, more information can be learned from the data. This trend continues until the hypothesis space grows to a region that cannot be probed by the available data. This results in $D_{KL}$ rolling off at small $f_{min}$. In the faint limit, $f_{min} \to 0$, no information is gained. In contrast, the Bayesian evidence continues to grow at a decreasing rate. Hence the Bayesian update looses its predictive power since unjustified increase in the dimensionality of the hypothesis space cannot be penalized by a drop in the Bayesian evidence. The resulting information gain is given in Figure \ref{figr:info}.

The choice of $f_{min}$ for a particular run can be motivated by the associated information gain it offers. Note that the relative entropy cannot be negative since
\begin{align}
    D_{KL} \leq -\int \dd{\theta} P(\theta|D) \Bigg( \frac{P(\theta)}{P(\theta|D)} - 1\Bigg) \\
    D_{KL} \geq 1 - \langle P(\theta) \rangle \geq 0.
\end{align}

\section{Performance given under-sampled data}

\begin{figure*}[ht]
    \centering
    \begin{minipage}[b]{\linewidth}
        \includegraphics[width=\linewidth]{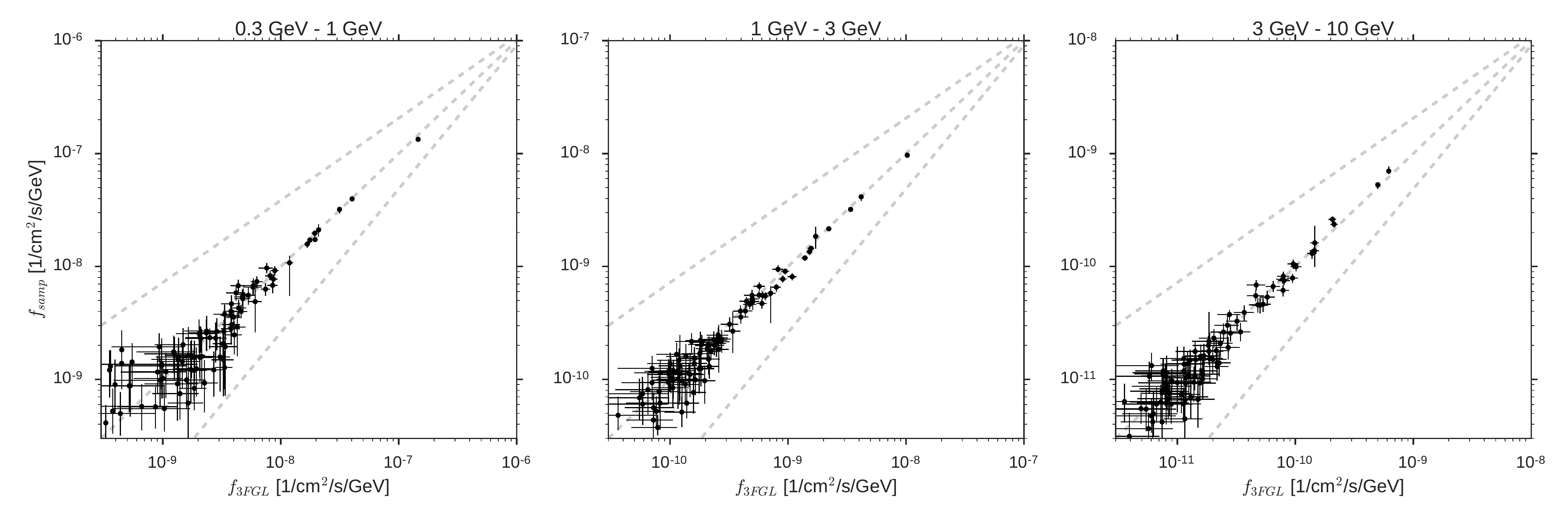}
    \end{minipage}
    \begin{minipage}[b]{\linewidth}
        \includegraphics[width=\linewidth]{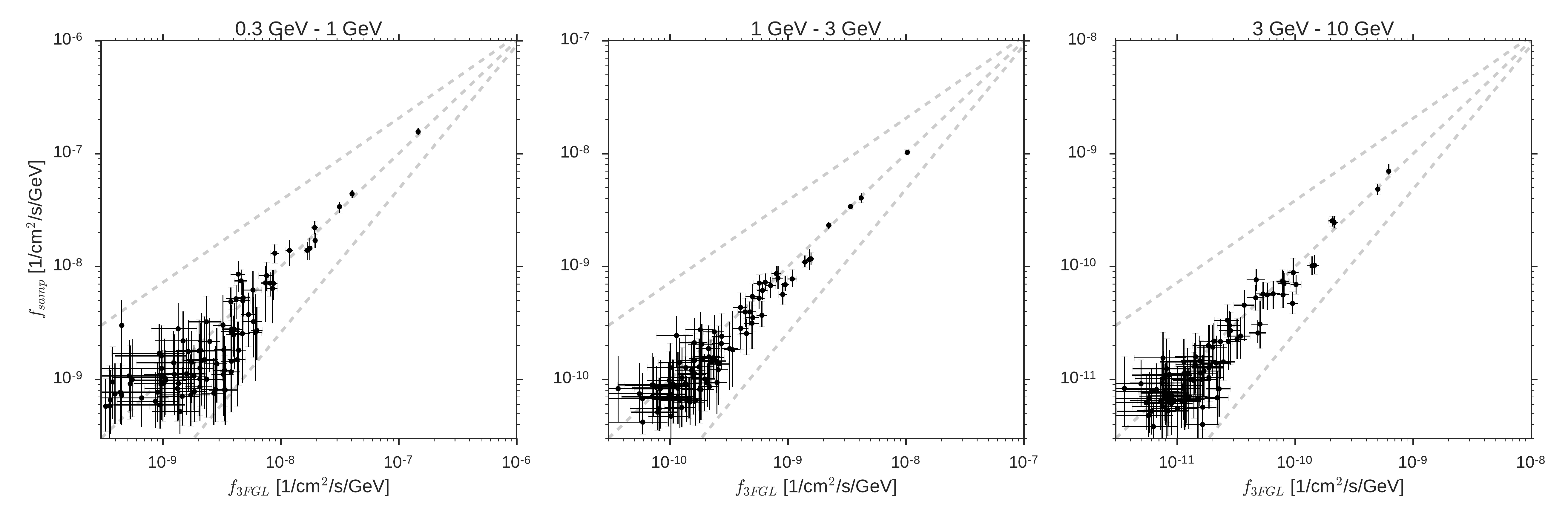}
    \end{minipage}
    \begin{minipage}[b]{\linewidth}
        \includegraphics[width=\linewidth]{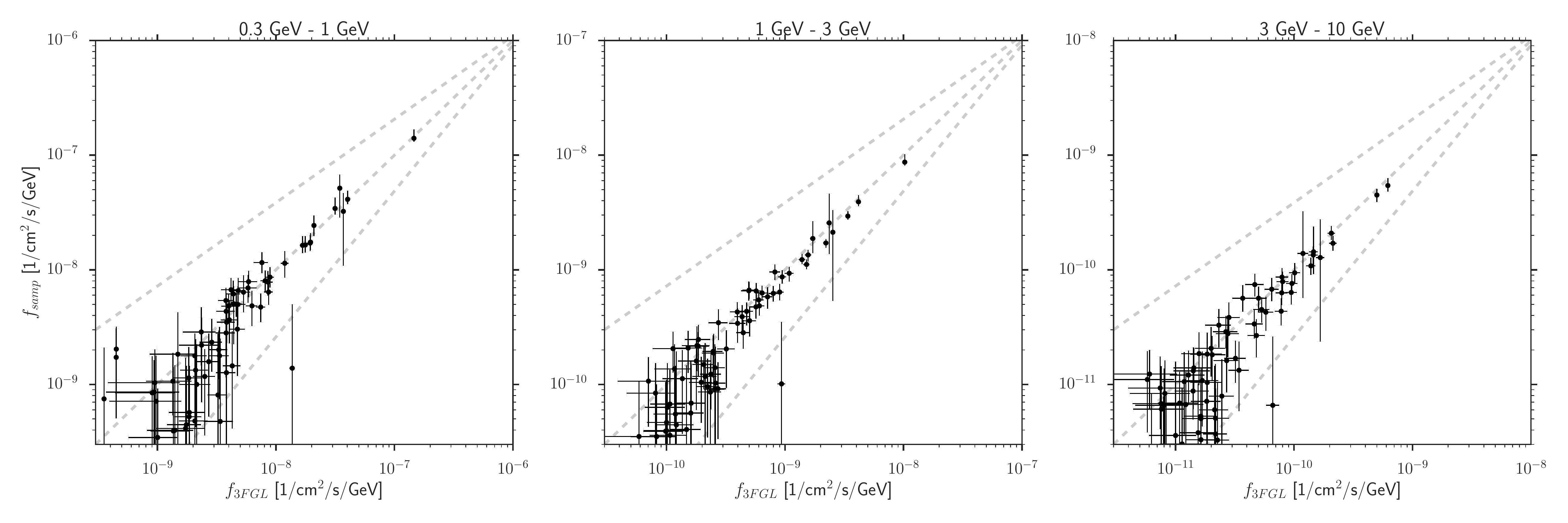}
    \end{minipage}
    \caption{Same as Figure \ref{figr:inptscatspec}, but using undersampled datasets. The fraction of data used is 50\% (top), 25\% (middle) and 10\% (bottom) of the available weekly files, respectively. The associations are still performed against the 3FGL catalog, which uses the full dataset ,i.e., weeks 9 through 217.}
    \label{figr:depl}
\end{figure*}

In order to assess the performance of the Bayesian approach, we further ask the question: How does probabilistic cataloging perform when fed by a down-sampled dataset? Towards this purpose we used the same Fermi-LAT dataset used above, but produced three more sky maps that contain 50\%, 25\% and 10\% of the dataset, respectively. Since most of the point sources in the ROI are variable over $\sim$ month time scale, using a contiguous set of weekly files would make the down-sampled datasets biased differently. Therefore we excluded every other $n^{th}$ weekly data file. In Figure \ref{figr:depl} we summarize the results of this exercise, which shows that probabilistic sampling gives reliable results at the faint end of the flux distribution even in the case of down-sampled, shot-noise dominated data. 

\section{Letting the PSF float}
\label{sect:psfnvari}
\begin{figure*}[ht]
    \centering
    \includegraphics[width=\linewidth]{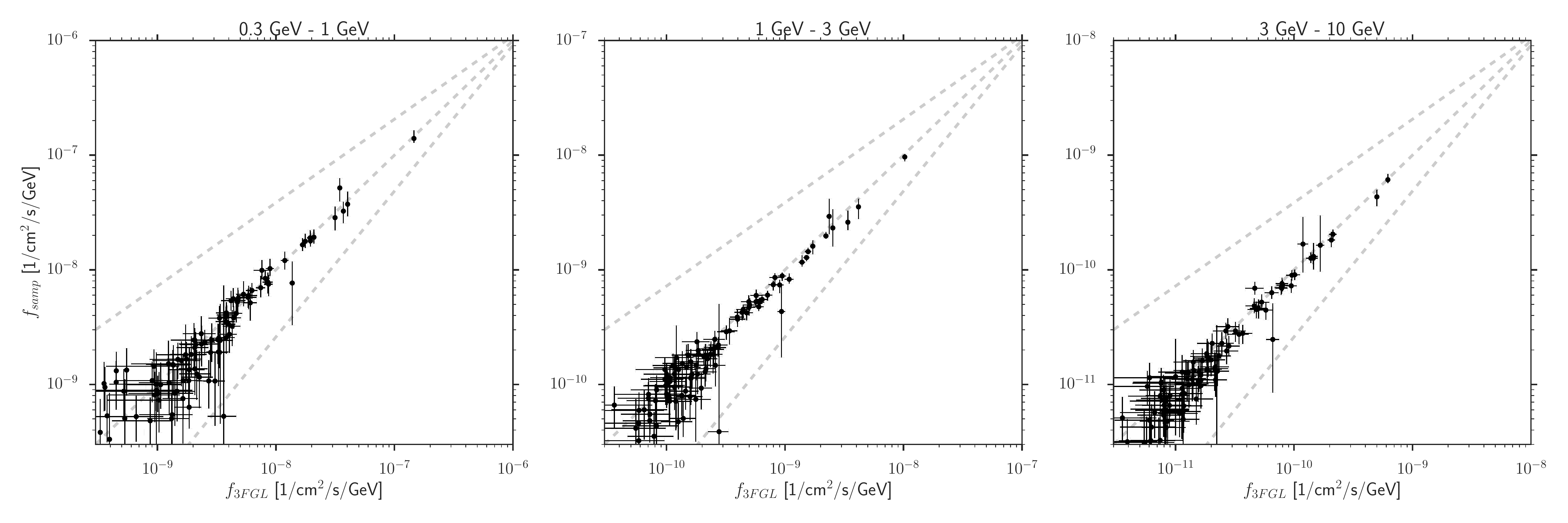}
    \caption{Same as Figure \ref{figr:inptscatspec}, but floating the PSF parameters.}
    \label{figr:inptscatspecpsfn}
\end{figure*}

Here we provide the association with the 3FGL obtained by floating the PSF parameters. Figure \ref{figr:inptscatspecpsfn} shows the correlation, which is weaker compared to the nominal results shown in Figure \ref{figr:inptscatspec}. This is expected, since the sampler is extracting as much information from the image as possible without the aid of a calibrator point source. In other words, our state of knowledge of the NGPC would be degraded by the amount shown, if we did not have strong priors about the radial profile of the PSF.

\section{Probabilistic catalog data format}
\label{sect:fits}
We make our results public in the form of an HDF5 file, which contains samples from the catalog space as well as relevant secondaries. The dataset can be accessed at the Harvard Dataverse, \url{https://dataverse.harvard.edu}. The contents of the HDF5 file are summarized in Table \ref{tabl:fits}, where an explanation is given for the image header-data unit contained in each extension.

\begin{table*}[ht]
    \caption{HDF5 file contents}
    \centering
    \begin{tabular}{c c c}
        Category & Extension Name & Explanation \\
        \hline
        Catalog samples & & \\
        & lgalpop0 & Galactic longitude [degree] \\
        & bgalpop0 & Galactic latitude [degree] \\
        & fluxpop0 & Flux density [1/cm$^2$/s/GeV] \\
        & sindpop0 & Spectral power-law index \\
        & meanpnts & mean number of point sources \\
        & fluxdistslop & slope of the flux distribution function \\
        & psfp & PSF parameters in each PSF class and energy bin \\
        & bacp & Normalization of the background emission templates in each energy bin \\
        \hline
    \end{tabular}
    \label{tabl:fits}
\end{table*}

\newpage

\bibliography{refr} 

\newpage
\listofchanges
\end{document}